\documentclass{article}

\usepackage{amsthm,amsmath,amsfonts,amssymb}
\usepackage{natbib}

\usepackage[latin1]{inputenc}
\usepackage[english]{babel} 
\usepackage{subfigure}
\usepackage{setspace}
\usepackage{tabularx} 
\usepackage{ctable} 
\usepackage{enumerate}
\usepackage{epsfig}
\usepackage{graphics}
\usepackage{graphicx}
\usepackage{color}
\usepackage{booktabs} 
\usepackage{longtable}
\usepackage{url}

\usepackage[short,12hr]{datetime}
\usepackage{fullpage}
\usepackage{srcltx}
\usepackage{flushend}
\usepackage{mathptmx}

\allowdisplaybreaks[4] 
\newcommand{\barma}{${\beta}$ARMA}
\newcommand{\tablesize}{\fontsize{8}{10}\selectfont} 
\newcommand{\tablesizea}{\fontsize{6}{8}\selectfont} 

\title{{Bootstrap-based inferential improvements in beta autoregressive moving average model}}

\author{
Bruna Gregory Palm\thanks{B.~G.~Palm 
is with the
Departamento de Estat\'istica,
Universidade Federal de Pernambuco, PE, Brazil,
E-mail: brunagpalm@gmail.com}
\and
F\'abio M. Bayer\thanks{F.~M.~Bayer 
is with the
Departamento de Estat\'istica and LACESM,
Universidade Federal de Santa Maria, RS, Brazil,
E-mail: bayer@ufsm.br}
}
\date{}

\begin{document}

\maketitle

\doublespacing

\abstract{
\noindent
We consider the issue of performing accurate small sample inference in beta autoregressive moving average model, which is useful for modeling and forecasting continuous variables that assumes values in the interval $(0,1)$. The inferences based on conditional maximum likelihood estimation have good asymptotic properties, but their performances in small samples may be poor. This way, we propose bootstrap bias corrections of the point estimators and different bootstrap strategies for confidence interval improvements. Our Monte Carlo simulations show that finite sample inference based on bootstrap corrections is much more reliable than the usual inferences. We also presented an empirical application.


\section{Introduction}

Generally, autoregressive integrated moving average models (ARIMA)~\citep{Box2008} are used for modeling and forecasting of variables over time.
However, these models become inappropriate when it is 
not reasonable to assume normality to the variable of interest
$y$, as occurs with variables of type rates and proportions~\citep{Ferrari2004}, {considering that these models does not take into account the bounded nature of the data.}
In such cases, the occurrences of $y$ belong to the continuous interval ($0,1$). Some examples of variables in the standard unit interval ($0,1$) are: relative air humidity, proportion of defective items, percentage of stored energy, proportion of patients, mortality rate, etc. {The use of ARIMA models
can lead to predicted values outside the unit interval in which the variable is defined~\citep{Cribari2010}. For example, forecasts of relative air humidity can reach values greater than 100\% or mortality rates may have predicted values lower than 0\%.}

An alternative to adequately model the data would be to use transformations of the variable of interest, but this approach has certain limitations. In this case, the results would be interpreted in terms of the transformed variable and not in terms of the average of the variable of interest. Moreover, variables 
{such as rates and proportions are typically asymmetrically distributed}, 
leading to distorted inferential results in models assuming normality of the data \citep{Cribari2010,Pinheiro2011}. For those situations in which it is desired to model over time a continuous variable in the interval ($0,1$), it was proposed the beta autoregressive moving average model (\barma) \citep{Rocha2009}. In this model, as well as in the beta regression model \citep{Ferrari2004}, we assume that the variable of {interest follows the beta distribution}. The beta probability density function is very flexible and, unlike normal density, accommodates distributions that are symmetrical, asymmetrical, ``J shaped'', inverted ``J shaped'', among others. Figure~\ref{f:fig} shows several forms of beta density shown in Equation~\eqref{E:den}, considering different parameter values of mean ($\mu$) and precision ($\phi$) that index it. 

\begin{figure}[h] 
\begin{center} 
\subfigure[$\;\phi=20$]
{\label{f:densitys_10}\includegraphics[width=0.4\textwidth]{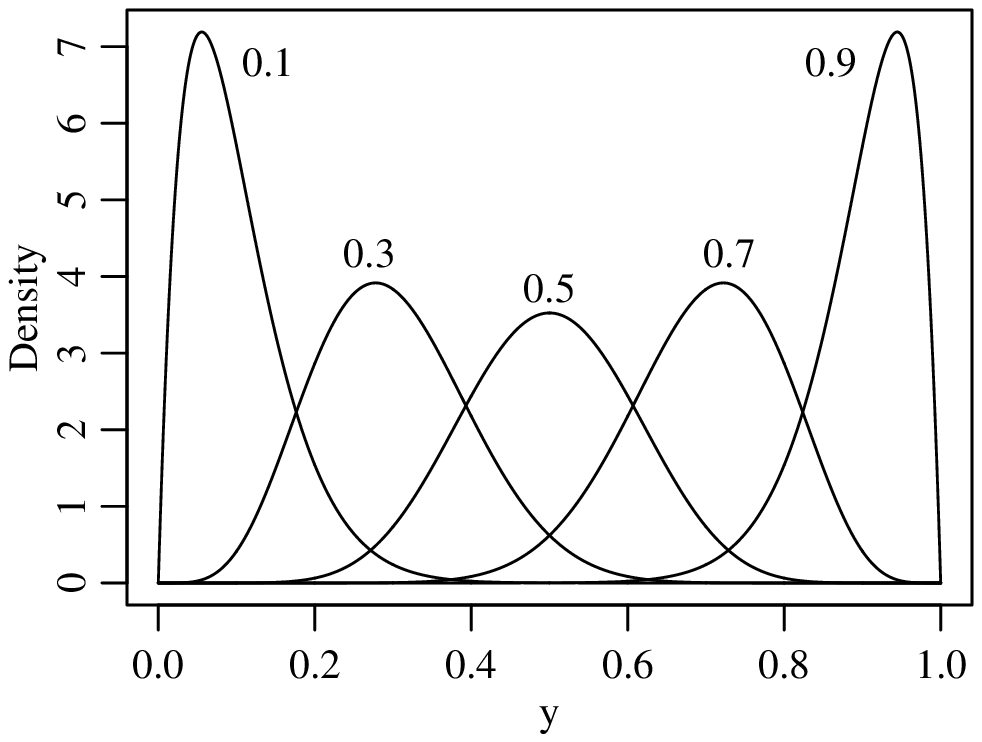}}
\subfigure[$\;\phi=120$]
{\label{f:densitys_90}\includegraphics[width=0.4\textwidth] {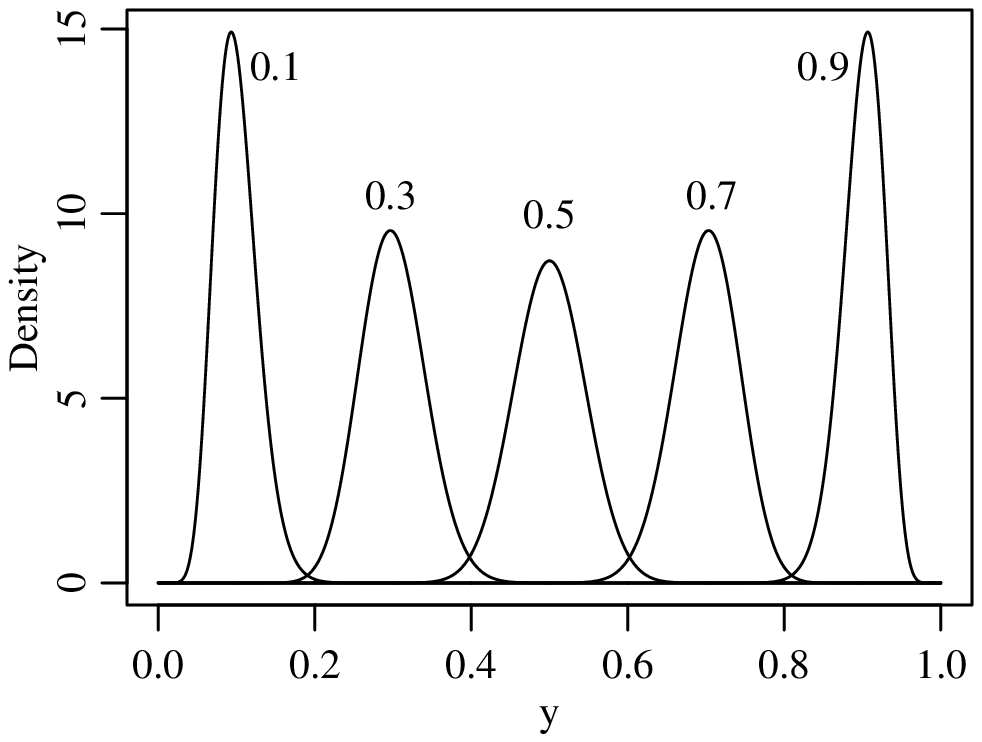}}
\caption{
Beta probability density functions with varying parameters 
$\mu=0.10$, $0.30$, $0.50$, $0.70$, $0.90$ for two values of $\phi$.
}\label{f:fig} 
\end{center} 
\end{figure} 

The specific inferences about the {\barma}~model parameters are based on conditional maximum likelihood estimation (MLE) and the interval inferences are based on the asymptotic properties of these estimators. These inferential procedures are also considered in ARIMA models, in generalized linear models (GLM)~\citep{McCullagh1989} and in beta regression~\citep{Ferrari2004}. MLE has good asymptotic properties, however, in small samples these approximations can be poor, generating distorted inferential results. Such distortions can occur both with respect to the bias of point estimators and in terms of coverage rate of confidence intervals. The biases of the MLE are on the order of $n^{-1}${~\citep{Cordeiro1994}}, where $n$ is the sample size. Thus, in samples of moderate size bias can become problematic~\citep{CordeiroCribari2014}. In this sense, inferential corrections for small samples become important research topics. 
Details on inferences in large samples based on likelihood are discussed in~\cite{Pawitan2001}. 

This paper proposes corrections for point and interval estimators in the \barma~model.  In particular, the bootstrap method~\citep{Efron1979} is considered to obtain bias-adjusted estimators and different strategies for improvement of the confidence intervals. 
The improved inference methods are also considered in an application to forecast a real data of percentage of energy stored. 
Such corrections become crucial for generating predicted values closer to nominal values, as seen in \cite{Kim2003}. Similar works (but in other classes of models) are verified in the literature. 
An extensive discussion on applications of the bootstrap method in time series models is presented in \cite{Berkowitz2000}. 
Bias correction by the bootstrap method in small samples in autoregressive models are studied in~\cite{Inoue2002} and~\cite{Kim2003}. Different bootstrap methods in time series are studied in \cite{Hardle2003} and evaluated in~\cite{Politis2003}. 
Bootstrap point and interval corrections in the beta regression model are discussed in~\cite{Ospina2006}, compared to analytical adjustments. 
{\cite{Cordeiro1994} obtained analytical bias corrections to the MLE parameters in the ARMA models.}
\cite{Franco2007} consider different bootstrap approaches and bootstrap confidence intervals to improve the inferences about the memory parameter in fractionally autoregressive moving average model.

The organization of this paper is arranged as follows. Section~\ref{s:barma} introduces the $\beta$ARMA model, as well as link function, conditional log-likelihood function and large sample inferential details. Section~\ref{s:corr} presents the inferential improvements in small samples by bootstrap methods. Section~\ref{s:num} describes the experiment of Monte Carlo simulation for finite sample size and presents the main numerical results and their discussion. Section ~\ref{s:aplicacao}  shows and discusses an application to real data, in order to compare the predictive performance of the models with corrected  and uncorrected estimators. At last, Section~\ref{s:con} presents the conclusion of the work.

\section{The beta autoregressive moving average model}\label{s:barma} 

{The \barma~model} was proposed by~\cite{Rocha2009} and can be defined as follow. Let $y = (y_1,\ldots,y_n)^\top$  be a vector of $n$ random variables, where each $y_t$, $t=1,\ldots,n$, has conditional distribution, given by a set of previous information $\mathcal{F}_{t-1}$, following beta distribution with mean parameters $\mu_t$ and precision $\phi$. The conditional density of $y_t$, given $\mathcal{F}_{t-1}$, is given by: 
\begin{equation}\label{E:den} 
f(y_t\mid\mathcal{F}_{t-1})=\frac{\Gamma(\phi)}{\Gamma(\mu_t\phi)\Gamma((1-\mu_t)\phi)}y_t^{\mu_t\phi-1}(1-y_t)^{(1-\mu_t)\phi-1},  
\end{equation} 
where $\Gamma(\cdot)$ is the gamma function, 
{$0<\mu_t <1$, and $\phi > 0$.}

The conditional mean and conditional variance of $y_t$ are respectively given by: \[ E(y_t\mid \mathcal{F}_{t-1})= \mu_t, \] \[ var(y_t \mid \mathcal{F}_{t-1})= V(\mu_t)/(1+\phi), \] where $V(\mu_t)=\mu_t(1-\mu_t)$ is denoted by variance function and $\phi$ can be interpreted as a {precision parameter. We note that the reciprocal
of the precision parameter can be viewed as a dispersion parameter.}

The \barma$(p,q)$ model is defined by the following structure: 
\begin{equation}\label{E:modelo} 
g(\mu_t)=\alpha+\sum\limits_{i=1}^{p}\varphi_i g(y_{t-i})+\sum\limits_{j=1}^{q}\theta_jr_{t-j}, 
\end{equation} 
where $\alpha \in\Re $ is a constant, $\varphi's$ and $\theta's$ are, respectively, the autoregressive and moving average parameters, 
$r_t$  
is the moving average error term 
and $g(\cdot)$ is a strictly monotone link function and twice differentiable where $g:(0,1)\rightarrow\Re $, as in the beta regression model~\citep{Ferrari2004}, and $p$ and $q$ are the orders of the model. The usual link functions for models assuming beta distribution are the logit, probit and complementary log-log. It should be noted that the seminal \barma~model, proposed in~\cite{Rocha2009}, also considers a term which accommodates covariates in the model, similar to the regression model. 
In this paper we consider $r_t =y_t-\mu_t$ for the error term.  
In this case, the model input $ r_t $ is in scale of $y$, and the model output is in the scale of $g(y_t)$. 
This way, the stationary rules applied to ARMA models can not be valid. 
It can happens because the transfer function of the system, 
that transforms $ r_t $ into $ y_t $, by a dynamic nonlinear 
relationship, 
can lead to uncontrollable system
~\citep{Box2008}.

{Parameter estimation can be performed using the maximum likelihood method.} 
Let the parameter vector be $\gamma=(\alpha,\varphi^\top,\theta^\top,\phi)^\top$, where $\varphi=(\varphi_1,\ldots,\varphi_p)^\top$ and $\theta=(\theta_1,\ldots,\theta_q)^\top$. The MLE are obtained by maximizing the logarithm of the conditional likelihood function. 
The log-likelihood function for the parameter vector $\gamma$ conditional to the $m$ preliminary observations, where $m=\max(p,q)$, can be defined as: 
\begin{equation}\label{E:vero} 
\ell=\ell(\gamma;y)=\sum\limits_{t=m+1}^{n} \log f(y_t\mid\mathcal{F}_{t-1})= \sum\limits_{t=m+1}^{n}\ell_t(\mu_t,\phi),
\end{equation} 
where $\ell_t(\mu_t,\phi)=\log\Gamma(\phi)-\log\Gamma(\mu_t\phi)-\log\Gamma((1-\mu_t)\phi) + (\mu_t \phi-1)\log y_t+\lbrace (1-\mu_t)\phi -1\rbrace \log (1-y_t)$. 

For the maximization of the function in \eqref{E:vero}, the use of nonlinear optimization algorithms is required. The computational implementation of this work was performed using the quasi-Newton optimization algorithm known as BFGS \citep{press} with analytic first derivatives. 
{
The optimization algorithm requires initial values. 
The 
starting values of the constant ($\alpha$) and the autoregressive ($\varphi$)  parameters 
were selected from an ordinary least squares estimate from a linear regression,  
where  $Y=(g(y_{m+1}), g(y_{m+2}), \ldots, g(y_{n}))^\top$ are the responses and the covariates matrix is given by
\begin{align*}
X=
\begin{bmatrix}
1& g(y_{m}) & g(y_{m-1}) & \cdots & g(y_{m-p}) \\
1& g(y_{m+1}) & g(y_{m}) & \cdots & g(y_{m-p+1}) \\
\vdots & \vdots & \vdots & \ddots & \vdots \\
1& g(y_{n-1}) & g(y_{n-2}) & \cdots & g(y_{n-p}) \\
\end{bmatrix}. 
\end{align*}
For the parameter $\theta$,
the starting values are setted equal to zero. 
The initial value of $\phi$
is considered in the same way as in the beta regression~\citep{Ferrari2004}.}
For more theoretical details regarding large sample inferences and matrix expressions to the score vector and the Fisher information matrix $(K(\gamma))$, see \cite{Rocha2009}.

For inferences in large samples, it is necessary to know the matrix of asymptotic variances and covariances of MLE, given by the inverse of $K({\gamma})$. The joint Fisher information matrix for $\alpha$, $\phi$, $\varphi$, $\theta$ is not a diagonal block matrix, thus the parameters are not orthogonal. This feature makes the \barma~model different of the dynamic models based on GLM~\citep{Rocha2009,garma} and on ARIMA models. Under usual regularity conditions and for large sample sizes, the MLE have $k$-multivariate normal distribution, being $k=p+q+2$, defined by: 
\[ \left(\begin{array}{ll} \widehat{\alpha}\\ \widehat{\phi}\\ \widehat{\varphi}\\ \widehat{\theta} \end{array} \right ) \sim \mathcal{N}_{(k)} \left(\begin{array}{ll} \left(\begin{array}{ll} \alpha \\ \phi \\ \varphi \\ \theta
\end{array} \right ), {K^{-1}(\gamma)} \end{array} \right ) , \] 
where $\widehat{\alpha},$ $\widehat{\phi},$ $\widehat{\varphi}$ and $\widehat{\theta}$ are the maximum likelihood estimators of $\alpha,$ $\phi,$ $\varphi$ and $\theta$, respectively. 

The MLE $\widehat{\gamma}$ and $K(\widehat{\gamma})$ are consistent estimator of $\gamma$ and $K(\gamma)$, respectively. Assuming that $J(\gamma)=\lim_{n\rightarrow \infty} K(\gamma)/n$ exists and it is nonsingular, we have $\sqrt{n}(\widehat{\gamma}-\gamma) \stackrel{\mathcal{D}}{\longrightarrow} \mathcal{N}_{k} (0,J(\gamma)^{-1})$, with $\stackrel{\mathcal{D}}{\longrightarrow}$ denoting convergence in distribution. Thus, if $\gamma_r$ denotes the $r$-th component of $\gamma$, it follows that: $(\widehat{\gamma}_r - \gamma_r) \lbrace K_{rr}^{-1}(\widehat{\gamma})\rbrace^{1/2} \stackrel{\mathcal{D}}{\longrightarrow} \mathcal{N}(0,1) $, where  $K_{rr}^{-1}(\widehat{\gamma})$ is the $r$-th element of the diagonal of $K ^{-1}(\widehat{\gamma})$. 
If $0 < \alpha < 1/2$ and $z_{\delta}$ represents the $\delta$ quantile of the distribution 
$\mathcal{N} (0,1)$, we have the following asymptotic confidence intervals for $\gamma_r$ with confidence $100(1-\alpha)\%$, for $r=1,\ldots,k$: \begin{equation}\label{E:intervaloass} 
[\widehat{\gamma}_r - z_{1-\alpha/2} (K^{-1}_{rr}(\widehat{\gamma}))^{1/2};\widehat{\gamma}_r + z_{1-\alpha/2} (K_{rr}^{-1}(\widehat{\gamma}))^{1/2}]. 
\end{equation} 
These intervals will be considered in Section~\ref{s:intervalos}. These approximate confidence intervals can have distortions in small samples, because the asymptotic pivotal quantities used in their construction may have asymmetric distribution and nonzero mean \citep{Ospina2006}. In addition, 
these confidence intervals may include values outside the parameter space~\citep{Ospina2006,CordeiroCribari2014}. Further details on the asymptotic confidence intervals can be found in~\cite{Davison1997} and \cite{Efron1994}.

In order to produce forecasts, the MLE of ${\gamma}$, $\widehat{\gamma}$, must be used to obtain estimates
for $\mu_t$, $\widehat{\mu}_t$ \citep{Rocha2009}. 
This way, the mean response estimate at $n+h$, where $h=1,2,\ldots$, is given by
\begin{align}\label{e:futuros}
\widehat{\mu}_{n+h}=g^{-1} \left(  \widehat{\alpha}+\sum\limits_{i=1}^{p}\widehat{\varphi}_i  \left[ g(y_{n+h-i}) \right]  + \sum\limits_{j=1}^{q}\widehat{\theta}_j \left[ r_{n+h-j} \right] \right),
\end{align}
where 
\begin{align*}
\left[ g(y_{n+h-i}) \right] & =\left\{\begin{array}{rc}
g(\widehat{\mu}_{n+h-i}),&\textrm{if}\;\;\; i< h,\\
g(y_{n+h-i}), &\textrm{if}\;\;\; i\geq h,
\end{array}\right.\\
\left[ r_{n+h-j} \right] &=\left\{\begin{array}{rc}
0,&\textrm{if}\;\;\; j< h,\\
\widehat{r}_{n+h-j}, &\textrm{if}\;\;\; j\geq h,
\end{array}\right.
\end{align*}
and $\widehat{r}_{t} = y_t - \widehat{\mu}_t$.

\section{Inferential improvements in small samples}\label{s:corr} 

In general, the MLE are biased to their true parametric values when the sample size is small. In practice the bias is often ignored, in justification of being negligible compared to the standard error of the MLEs. The standard deviation of the estimator is of $n^{-1/2}$ order, while the bias is of $n^{-1}$ order. However, in some models, the bias in small samples can be appreciable or have magnitude equal to the standard error of the estimator~\citep{CordeiroCribari2014,Davison1997}. 

Aiming to reduce the problem of MLE bias in small samples, \cite{Cox1968} proposed a very general analytical formula to determine the bias of order $O(n^{-1})$ of MLE in multi-parametric models. To determine it, we should know the inverse of the Fisher information matrix and cumulants of log-likelihood derivates up to third order with respect to the unknown parameters{~\citep{Cordeiro1994}}. From the determination of the bias, we can set the second order MLE by:
\begin{align*} 
\overline{\gamma}=\widehat{\gamma}-\widehat{B}(\widehat{\gamma}), 
\end{align*} 
where $\widehat{B}(\cdot)$ is the bias $B(\cdot)$ evaluated in $\widehat{\gamma}$. The bias of the corrected estimator will be of order $n^{-2}$, that is, \begin{align*} E(\overline{\gamma})=\gamma+O(n^{-2}). \end{align*} 

However, analytical derivation as shown in \cite{Cox1968} can be difficult to obtain, or even impossible to be determined to certain classes of models. The calculation of higher order moments and cumulants is rather complicated~\citep{CordeiroCribari2014}. In particular, in \barma~models where the parameters are not orthogonal, this analytical derivation is especially costly. In this sense, bootstrap corrections become good options for inferential improvements in small samples. In this approach, the bias estimation $B(\cdot)$ is numerically obtained through Monte Carlo simulations, bypassing analytical difficulties. 
Bootstrap bias-corrected estimators are discussed in more details in Section~\ref{s:corevies}. 

The bootstrap method is a computationally intensive method based on resampling, being useful for inferential corrections on small samples~\citep{Efron1979}. 
Basically, there are two possible bootstrap approaches: parametric and nonparametric. 
In the nonparametric method the pseudo samples are generated from the originally observed data. 
For the parametric method, 
a parametric model is fitted to the original data and pseudo samples from this fitted model are generated~\citep{Davison1997, Efron1994}.
{In what follows we shall use the parametric bootstrap.}

The parametric bootstrap method can be generalized as follows: \begin{enumerate} \item Suppose that $y=(y_1,\ldots,y_n)^\top$ is a random sample that follows a distribution $F$ with parametric vector ${\gamma}$ ; 

\item From the original sample, obtain the estimates $\widehat{\gamma}$ of ${\gamma}$; 

\item \label{I:3} {Generate $B$ size $n$ bootstrap samples $(y_1^\ast, \ldots , y_B^\ast)$ from $F(\widehat{\gamma})$;}

\item \label{I:4} {For each bootstrap sample $y_b^\ast$ compute $\widehat{\gamma}^\ast$;} 

\item Repeat steps~\ref{I:3} and~\ref{I:4} a  very large number $B$ of times, thus obtaining: $\widehat{\gamma}^{\ast 1},\ldots,\widehat{\gamma}^{\ast B}$; 

\item Use the estimates $\widehat{\gamma}^{\ast b}$, with $b=1,\ldots,B$, to calculate the desired quantities (mean, variance, confidence interval, etc). 

\end{enumerate} 

\subsection{Bias correction of point estimators}\label{s:corevies} 

The MLE are asymptotically not biased, however, their bias in small samples can be considerable \citep{Pawitan2001,Efron1994}. Through bootstrap method, we can estimate the bias of a point estimator. Once we have a good estimate of the bias of the estimator, we can build bias-corrected point estimators. The bias of the estimator $\widehat{\gamma}$ can be expressed as: \[ B(\widehat{\gamma})=E[\widehat{\gamma}]-\gamma. \] 
Using the steps of the bootstrap method previously presented, a bootstrap estimate of the bias can be obtained by \[ \widehat{B}_{boot}(\widehat{\gamma})=\bar{\gamma}^{\ast} -\widehat{\gamma}, \] where $\bar{\gamma}^{\ast}=\frac{1}{B} \sum \limits _{b=1}^{B} \widehat{\gamma}^{\ast b}$. Thus, we can obtain a corrected second order estimator~\citep{Efron1979,Davison1997}: 
\begin{equation} \label{E:corr} \begin{array}{rcl} \overline{\gamma}=\widehat{\gamma}-\widehat{B}_{boot}(\widehat{\gamma})= \widehat{\gamma}-(\bar{\gamma}^{\ast} -\widehat{\gamma})=2\widehat{\gamma}-\bar{\gamma}^\ast. \end{array} 
\end{equation} 
This estimator has the same asymptotic properties as the usual MLE, but has less bias in small samples~\citep{Efron1994}. 

\subsection{Corrected confidence intervals}\label{s:intervalos} 

The general form for confidence intervals (CI) for $\gamma$ is: 
\begin{align*}
P_F[L\leq\gamma \leq U]\approx 1-\alpha,
\end{align*} 
where $L$ and $U$ are the lower and upper bounds of the confidence limits, respectively and {$\alpha \in (0, 0.5)$}. 
In likelihood inferences, this interval requires large samples to guarantee the validity of the asymptotic approximations. In small samples their effectiveness can be seriously compromised~\citep{Efron1979,Davison1997}. 

An alternative to the construction of adequate confidence intervals in small samples, free of analytical complexities, is the bootstrap method. One advantage is the independence of the central limit theorem, because their precision measurements are obtained directly from the data. Bootstrap intervals are approximate as standard confidence intervals, despite the fact that better approximations can be offered~\citep{Efron1994}. 

Confidence intervals for $\gamma$ can be obtained in several ways. One of the most common ways is, as introduced earlier in Equation~\eqref{E:intervaloass}, given by: \begin{equation} \label{E:intervalo1} [\widehat{\gamma}-z_{(1-\alpha /2)}\widehat{ep}(\widehat{\gamma});\widehat{\gamma}+z_{(1-\alpha /2)}\widehat{ep}(\widehat{\gamma})], \end{equation} where $\widehat{ep}(\widehat{\gamma})$ is the estimate of the standard error of $\widehat{\gamma}$, in which ${\rm diag}\{{(K^{-1}(\widehat{\gamma})})^{1/2}\}$ is usually used as approximation for $\widehat{ep}(\widehat{\gamma})$, when $\widehat{\gamma}$ is MLE. 

Intervals in the way of Equations \eqref{E:intervaloass} and \eqref{E:intervalo1} are approximate, with coverage probability not exactly equal to $1-\alpha$, as desired~\citep{Efron1979}. Using bootstrap, we can have a better estimate of the standard error of the estimator $\widehat{\gamma}$, given by: 
\begin{equation} \label{E:ep} 
\widehat{ep}_{boot}(\widehat{\gamma})={\sqrt{\frac{\sum \limits _{b=1}^B(\gamma^{\ast b}-\bar{\gamma}^{\ast})^2}{B-1}}}. 
\end{equation} 
The standard bootstrap confidence interval $(CI_{boot})$ is obtained through the bootstrap estimate of the standard error, given by~\eqref{E:ep}, with coverage probability of approximately $1-\alpha$, given by: \begin{equation}\label{E:intervaloboot} [\widehat{\gamma}- z_{(1-\alpha /2)} \widehat{ep}_{boot}(\widehat{\gamma});\widehat{\gamma}+ z_{(1-\alpha /2)} \widehat{ep}_{boot}(\widehat{\gamma})]. \end{equation} The major advantage of this method is its algebraic simplicity for finding a CI for $\gamma$. A desired property of the intervals is the preservation of the range, which is not always satisfied in the standard bootstrap interval.

The bootstrap-$t$ interval $(CI_{t})$~\citep{Efron1994}, also known as pivotal method is a generalization of the $t$-Student method, and is usually applied in location statistics as sampling mean, median, or sampling percentile. 
Let the $\alpha$-th percentile of the $t$-distribution be denoted by  $t_{(\alpha)}$, thus the $CI_{t}$ is given by: 
\[ [\widehat{\gamma} - {t}_{(1-\alpha/2)} \widehat{ep}_{boot}(\widehat{\gamma});\widehat{\gamma} + {t}_{(1-\alpha/2)} \widehat{ep}_{boot}(\widehat{\gamma})]. \] 

The bootstrap percentile interval $(CI_{p})$~\citep{Efron1994} has the property of invariance to monotonic transformations. It is constructed from a finite number $B$ of  bootstrap replications of the estimator of the parameter of interest. Thus, the percentile confidence interval is given by: \[[\widehat{\gamma}^{\ast B}_{(\alpha /2)};\widehat{\gamma}^{\ast B}_{(1-\alpha /2)}], \] being $\widehat{\gamma}^{\ast B}_{(\alpha /2)}$ the $(100 \cdot \alpha /2)$-th percentile of the resamplings $\widehat{\gamma}^{\ast b}$, that is, the $(B\cdot \alpha /2)$-th value of an ordering of $B$ replications of $\widehat{\gamma}^{\ast b}$. 

This work still considers another corrected confidence interval. The CI based on point unbiased estimator $(CI_{u})$ obtained through of the confidence interval given in~\eqref{E:intervaloboot} replacing the MLE by their bias-corrected versions given in~\eqref{E:corr}. Thus, the $CI_{u}$ is defined as: 
\[ [\overline{\gamma} - z_{(1-\alpha /2 )}\widehat{ep}_{boot}(\widehat{\gamma}); \overline{\gamma} + z_{(1-\alpha /2 )}\widehat{ep}_{boot}(\widehat{\gamma})].\] 

The following Section evaluates the finite sample performances of different confidence intervals introduced in this Section, when used to make inferences about the parameters of the \barma~model.

\section{Numerical evaluation} \label{s:num}

The evaluation of the point and interval estimators, corrected and uncorrected, 
of the \barma~model parameters was performed through Monte Carlo simulations. 
The computational implementation was developed using the {\tt R} programming language 
~\citep{R2012}. 
The number of Monte Carlo and bootstrap replications were set equal to
$1,000$. The sample sizes considered were $n =20, \, 30, \, 50, \, 100$. 

The numerical results presented are based on the \barma~model with 
the mean structure
given by the \eqref{E:modelo} and logit link function; i.e. $\operatorname{logit}(\mu)=\log(\frac{\mu}{1-\mu})$. For the parameter values were considered different scenarios, namely: \begin{itemize} 
\item $\beta$AR$(1)$ with $\alpha=1$, $\varphi_1=-0.5$ and $\phi=20$; 
\item  $\beta$AR$(1)$ with $\alpha=1$, $\varphi_1=-0.5$ and $\phi=120$;
\item $\beta$MA$(1)$ with $\alpha=-1$, $\theta_1=1$ and $\phi=20$; 
\item $\beta$MA$(1)$ with $\alpha=1$, $\theta_1=-0.5$ and $\phi=120$; 
\item \barma$(1,1)$ with $\alpha=-0.5$, $\varphi_1=0.5$, $\theta_1=1$ and $\phi=20$;
\item \barma$(1,1)$ with $\alpha=1$, $\varphi_1=0.5$, $\theta_1=-1.5$ and $\phi=120$.

\end{itemize} For brevity, we will present the results of the scenarios with $\phi=20$  in this section, due to similarities in the results.
The results with $\phi=120$ can be found in the Appendix.

In order to numerically evaluate the point estimators, it is necessary to use some measurements. From the $1,000$ Monte Carlo replications of the maximum likelihood estimators, usual and corrected, we calculate mean, bias, percentage relative bias (RB), standard error (SE) and mean square error (MSE). A graphical analysis of the behavior of the RB is performed through the graph of total relative bias, defined in~\cite{Cribari2003} as the sum of the absolute values of the individual relative biases. Thus, the total relative bias is an aggregate measure of the bias of the parameters estimates.

Table~\ref{T:bar201} presents the results of numerical evaluation of point estimators of the parameters of $\beta$AR$(1)$ model. It is observed that the usual MLE $(\widehat{\alpha},\widehat{\varphi}_1, \widehat{\phi})$ of the $\beta$AR$(1)$ model parameters are considerably more biased than their corrected versions via bootstrap $(\overline{\alpha},\overline{\varphi}_1, \overline{\phi})$. Is is also noticeable that the non corrected estimators of $\phi$ present themselves more biased than the estimators of the autoregressive structure in all sample sizes. For $n=30$ we noted relative biases for $\widehat{\phi}$ and $\bar{\phi}$ equal to 14.882 and -0.935, respectively. That is, the uncorrected estimator is about 15 times {more biased than} the proposed corrected estimator. Moreover, as expected, by the asymptotic properties of the MLE, the bias of all the estimators decrease as the sample size increases. We can verify graphically from Figure \ref{f:viesbar} that the total relative bias is considerably smaller in the corrected estimators and converges faster to zero. 
Regarding the MSE, we verify that it decreases as the sample size increases in all estimators, which is a numerical indicative of the consistency of the estimators.

\begin{table}[t]
\caption{Results of the Monte Carlo simulation for point estimation in $\beta$AR$(1)$ model.} 
\label{T:bar201} 
\tablesize
\begin{center} 
\begin{tabular}{rrrrrrr} 
\hline
Measures 	&	$\widehat{\alpha}$ 	&	$\overline{\alpha}$  	&	 $\widehat{\varphi}_{1}$ 	&	 $\overline{\varphi}_{1}$  	&	$\widehat{\phi}$ 	&	$\overline{\phi}$		\\                      
\hline                 
\multicolumn{7}{c}{$n=20$}\\ 
\hline	
Mean &  $0.991$ & $1.000$ & $-0.476$ & $-0.494$ & $24.135$ & $19.122$ \\
Bias  & $-0.009$ & $0.000$ & $0.024$ & $0.006$ & $4.135$  & $-0.878$ \\
RB   &  $-0.896$ & $-0.027$ & $-4.735$ & $-1.244$ & $20.675$ & $-4.392$ \\
SE  &  $0.176$ &  $0.192$ & $0.194$ & $0.221$ & $8.635$ & $7.389$ \\
MSE   &  $0.031$  &  $0.037$ &  $0.038$ &  $0.049$ & $91.658$ &   $55.370$ \\

\hline
\multicolumn{7}{c}{$n=30$}\\                           
\hline  
Mean &  $0.988$  &    $0.995$  & $-0.476$  &  $-0.490$ & $22.976$ & $19.813$\\
Bias  & $-0.012$  & $-0.005$  & $0.024$  &   $0.010$ & $2.976$  & $ -0.187$ \\
RB   &  $-1.189$  &   $-0.487$ & $-4.812$  &  $-2.066$ & $14.882$  & $-0.935$\\
SE  &  $0.140$  &  $0.149$  & $0.151$  &  $0.165$ & $6.906$ &  $6.275$\\
MSE   &  $0.020$  & $0.022$  & $0.023$ &  $0.027$ & $56.558$ & $39.410$ \\

 \hline
\multicolumn{7}{c}{$n=50$}\\                           
\hline
Mean &  $0.993$   &   $0.999$ & $-0.485$ &   $-0.495$ & $21.807$ &   $20.029$ \\
Bias &   $-0.007$  &   $-0.001$  & $0.015$  &   $0.005$ & $1.807$  &   $0.029$ \\
RB  &   $-0.670$   &  $-0.126$  & $-3.083$  &  $-1.048$ & $9.037$  &   $0.146$ \\
SE    &  $0.107$  &    $0.111$  & $0.118$  &   $0.125$ & $4.615$  &   $4.319$ \\
MSE   &  $0.011$   &   $0.012$  & $0.014$   &  $0.016$ & $24.563$  &  $18.653$ \\

 \hline
 \multicolumn{7}{c}{$n=100$}\\                           
\hline
Mean  & $0.996$  &  $0.999$ & $-0.496$  &  $-0.502$ & $20.605$ & $19.767$ \\
Bias & $-0.004$  &   $-0.001$  & $0.004$  &  $-0.002$ & $0.605$  &  $-0.233$ \\
RB   &  $-0.408$  &   $-0.059$ & $-0.800$ &    $0.435$ & $3.027$ &   $-1.164$ \\
SE &   $0.077$    &  $0.078$  & $0.082$  &   $0.085$ & $2.995$ &    $2.868$ \\
MSE   &  $0.006$  &    $0.006$  & $0.007$  &   $0.007$ & $9.334$ &  $8.278$\\

\hline
\end{tabular}	
\end{center}	
\end{table} 

\begin{table}[t]
\caption{
Results of the Monte Carlo simulation for point estimation in $\beta$MA$(1)$ model.
} 
\label{T:bma201} 
\tablesize
\begin{center} 
\begin{tabular}{rrrrrrr} 
\hline
Measures 	&	$\widehat{\alpha}$ 	&	$\overline{\alpha}$  	&	 $\widehat{\theta}_{1}$ 	&	 $\overline{\theta}_{1}$  	&	$\widehat{\phi}$ 	&	$\overline{\phi}$		\\                      
\hline                 
\multicolumn{7}{c}{$n=20$}\\ 
\hline	
Mean & $-1.008$  &   $-1.007$  &  $0.747$   &    $1.112$ &$23.843$ &   $20.752$\\
Bias  & $-0.008$   &  $-0.007$  & $-0.253$  &     $0.112$ & $3.843$   &  $0.752$\\
RB   &   $0.792$  &    $0.731$ & $-25.273$  &    $11.180$ &$19.215$  &   $3.762$\\
SE  &     $ 0.135$  &    $0.135$  &  $1.551$   &    $1.606$ & $7.800$   &  $8.046$\\
MSE  &   $0.018$   &   $0.018$  &  $2.468$  &     $2.593$ &$75.609$   & $65.306$\\

\hline
\multicolumn{7}{c}{$n=30$}\\                           
\hline  
Mean &  $-1.002$  &   $-1.000$  &  $0.791$  &     $0.990$ &$22.698$ &   $20.240$\\
Bias  & $-0.002$  &    $0.000$ &  $-0.209$  &    $-0.010$ & $2.698$ &    $0.240$\\
RB    &  $0.205$   &   $0.038$ & $-20.921$   &   $-1.000$ & $13.488$ &    $1.201$\\
SE  &   $0.107$  &    $0.107$  &  $1.078$  &     $1.080$ & $6.121$  &   $6.105$\\
MSE   &  $0.011$  &    $0.012$  &  $1.206$  &     $1.167$ &$44.747$ &   $37.333$\\

 \hline
\multicolumn{7}{c}{$n=50$}\\                           
\hline
Mean & $-1.005$  &   $-1.004$ &   $0.866$  &     $0.974$ &$21.270$ &   $19.840$\\
Bias  & $-0.005$   &  $-0.004$  & $-0.134$  &    $-0.026$ & $1.270$ &   $-0.160$\\
RB    &  $0.516$   &   $0.396$ & $-13.422$  &    $-2.647$  &$6.349 $  & $-0.798$\\
SE  &  $0.089$  &    $0.089$ &   $0.737$  &     $0.732$ & $4.210$ &    $4.124$\\
MSE   &  $0.008$ &     $0.008$  &  $0.562$    &   $0.537$ & $19.337$ &   $17.031$\\

 \hline
 \multicolumn{7}{c}{$n=100$}\\                           
\hline
Mean &  $-1.003$   &  $-1.002$  &  $0.948$   &    $0.992$ &$20.655$ &   $19.818$\\
Bias &   $-0.003$  &   $-0.002$  & $-0.052$   &   $-0.008$ & $0.655$  &  $-0.182$\\
RB    &  $0.284$   &   $0.179$ &  $-5.157$   &   $-0.803$  &$3.273$ &   $-0.910$\\
SE    &    $ 0.061$   &   $0.061$  &  $0.526$  &     $0.524$ & $2.910$  &   $2.811$\\
MSE   &  $0.004$   &   $0.004$  &  $0.280$   &    $0.274$ & $8.895$ &    $7.932$\\

\hline
\end{tabular}	
\end{center}	
\end{table}

\begin{table}[t] 
\caption{
Results of the Monte Carlo simulation for point estimation in \barma$(1,1)$ model.
} 
\label{T:barma201} 
\tablesize

\begin{center} 
\begin{tabular}{rrrrrrrrr} 
\hline
Measures 	&	$\widehat{\alpha}$ 	&	$\overline{\alpha}$  	&	$\widehat{\varphi}_{1}$ 	&	 $\overline{\varphi}_{1}$ &$\widehat{\theta}_{1}$ 	&	 $\overline{\theta}_{1}$  	&	$\widehat{\phi}$ 	&	$\overline{\phi}$		\\                      
\hline                 
\multicolumn{9}{c}{$n=20$}\\ 
\hline	
Mean &  $-0.687$  &   $-0.594$ &  $0.313$  &   $0.404$  &  $1.566$  &     $1.677$ &$22.350$&$20.195$\\
Bias  & $-0.187$  &   $-0.094$ & $-0.187$  &  $-0.096$  &  $0.566$  &     $0.677$ & $2.350$&$0.195$\\
RB   & $37.376$  &   $18.857$ &$-37.435$  & $-19.112$ &  $56.644$  &    $67.700$ &$11.751$&$0.976$\\
SE  & $0.370$   &   $0.500$ &  $0.315$  &   $0.469$  &  $2.049$ &      $2.949$ & $7.722$&$ 8.060$\\
MSE   &  $0.172$  &    $0.259$  & $0.134$  &   $0.229$  &  $4.517$  &     $9.155$ &$65.150$&$64.997$\\
\hline 
\multicolumn{9}{c}{$n=30$}\\                           
\hline  
Mean &   $-0.647$   &  $-0.554$ &  $0.367$  &   $0.457$  &  $1.453$ &      $1.324$ &$22.064$&$20.712$\\
Bias  & $-0.147$  &   $-0.054$ & $-0.133$  &  $-0.043$  &  $0.453$ &      $0.324$ & $2.064$&$0.712$\\
RB    &  $29.389$  &   $10.874$ &$-26.661$  &  $-8.577 $ & $45.287$ &     $32.384$ &$10.322$&$3.562$\\
SE  &  $0.316$  &    $0.400$  & $0.255$  &   $0.346$  &  $1.638$ &      $2.161$ & $6.326$&$6.587$\\
MSE   & $0.122$  &    $0.163$ &  $0.083$  &   $0.121$  &  $2.890$  &     $4.776$ &$44.278$&$43.898$\\       
\hline
\multicolumn{9}{c}{$n=50$}\\                           
\hline
Mean & $-0.598$  &   $-0.526$ &  $0.410$  &   $0.477$ &   $1.317$ &      $1.155$ &$21.310$&$20.333$\\
Bias  & $-0.098$ &    $-0.026$ & $-0.090$  &  $-0.023$  &  $0.317$  &     $0.155$ & $1.310$&$0.333$\\
RB    &  $19.574$  &    $5.247$ &$-17.930$ &   $-4.665$ &  $31.661$ &     $15.471$ & $6.552$&$1.666$\\
SE  & $0.222$   &   $0.254$ &  $0.187$  &   $0.225$ &   $1.145$ &      $1.328$ & $4.883$&$5.046$\\
MSE   &  $0.059$   &   $0.065$ &  $0.043$  &   $0.051$  &  $1.412$  &     $1.788$ &$25.563$&$25.568$\\
 
 \hline
      
 \multicolumn{9}{c}{$n=100$}\\                           
\hline
Mean & $-0.539$   &  $-0.500$ & $0.462$    & $0.500$ &   $1.087$  &     $1.022$ &$20.454$ &   $19.954$\\
Bias &   $-0.039$  &    $0.000$ &$-0.038$  &   $0.000$ &   $0.087$  &     $0.022$ & $0.454$   & $-0.046$\\
RB    & $7.898$  &   $-0.010$ &$-7.516$  &  $-0.083$  &  $8.736$  &     $2.166$ & $2.271$  &  $-0.229$\\
SE    &    $0.152$  &    $0.159$ & $0.128 $ &   $0.136$  &  $0.846$   &    $0.879$ & $3.251$  &   $3.229$\\
MSE   &  $0.025$   &   $0.025$ & $0.018$  &   $0.018$  &  $0.723$  &     $0.772$ &$10.776$  &  $10.427$\\
  
\hline

\end{tabular}	
\end{center}	
\end{table}

Analyzing Table~\ref{T:bma201} we also verify that the usual MLE of the $\beta$MA$(1)$ model are more biased than their corrected versions via bootstrap. The parameter $\phi$ shows itself percentually more biased than the estimators of the parameters of the mean structure in the smaller sample sizes in their usual versions. There was decreasing in the bias for all corrected estimators, reaching at almost null bias in the parameter $\alpha$ in sample sizes $n=20$ and $n=30$. The difference of the relative bias between the estimators $\widehat{\phi}$ and $\bar{\phi}$ is $6$ times in the sample size $n=20$, where $\widehat{\phi}$ presents RB equal to $19.215$ and $\bar{\phi}$ of $3.762$. This difference is even greater for $n=30$, where the difference between $\widehat{\theta}$ and $\bar{\theta}$ is $20$ times and between $\widehat{\phi}$ and $\bar{\phi}$ of $13$ times. Such estimates have relative bias of $-20.921$, $-1.000$, $13.488$ and $1.201$, respectively. The property of asymptotic non-bias of MLE is verified, since the bias decreases while sample size increases, as shown in Figure~\ref{f:viesbma}. It shows a decrease in the relative bias in the corrected estimators and the faster convergence for the parameter value.

Table~\ref{T:barma201} shows the results of the evaluation in the estimators of \barma$(1,1)$ model. Similarly to the cases $\beta$AR$(1)$ and $\beta$MA$(1)$, we observe lower values of bias for the corrected estimates compared to the usual ones. Corrected estimates were more accurate because they present lower value of bias. There is improvement through bias correction in almost all parameters, reaching an almost null value in sample size $n=100$ for $\bar{\alpha}$ and $\bar{\varphi_1}$. The largest difference of percentage bias is between $\widehat{\phi}$ and $\bar{\phi}$ for $n=20$, in which the uncorrected estimator has bias $11$ times higher than the corrected one. 
Figure \ref{f:viesbarma1} shows the lower values of the total relative bias in the corrected estimates, beside the convergence to zero is faster. In addition to that, it is also observed that the asymptotic property of unbiasedness of MLE is satisfied, because the bias decreases as the sample size increases.

We also verified the importance of correction in the MLE in the \barma~model.
The usual estimates were biased in different scenarios and sample sizes. The corrected estimators proved considerably less biased compared to the usual uncorrected. In general, the estimates performed better in the autoregressive estimator than in the part of the moving averages. Such fact was already discussed in \cite{Ansley1980} in ARMA models, where it is found, by simulation studies, that inferences about parameters of moving averages are poor. It is also noticeable that the estimator of $\phi$ proved itself biased in the usual statistics in different simulated models in almost all sample sizes. Such fact is also verified in the beta regression model, where the estimators of the precision $(\widehat{\phi})$ are percentually more biased than those of the mean structure, and the bootstrap corrections in the MLE of $\phi$ greatly reduce the bias in the corrected estimates \citep{Ospina2006}. Thus, we recommend the use of corrected statistics rather than classical MLE to obtain better point estimators. 

\begin{figure}[t] 
\begin{center} 
\subfigure[$\beta$AR$(1)$ model]
{\label{f:viesbar}\includegraphics[width=0.4\textwidth]{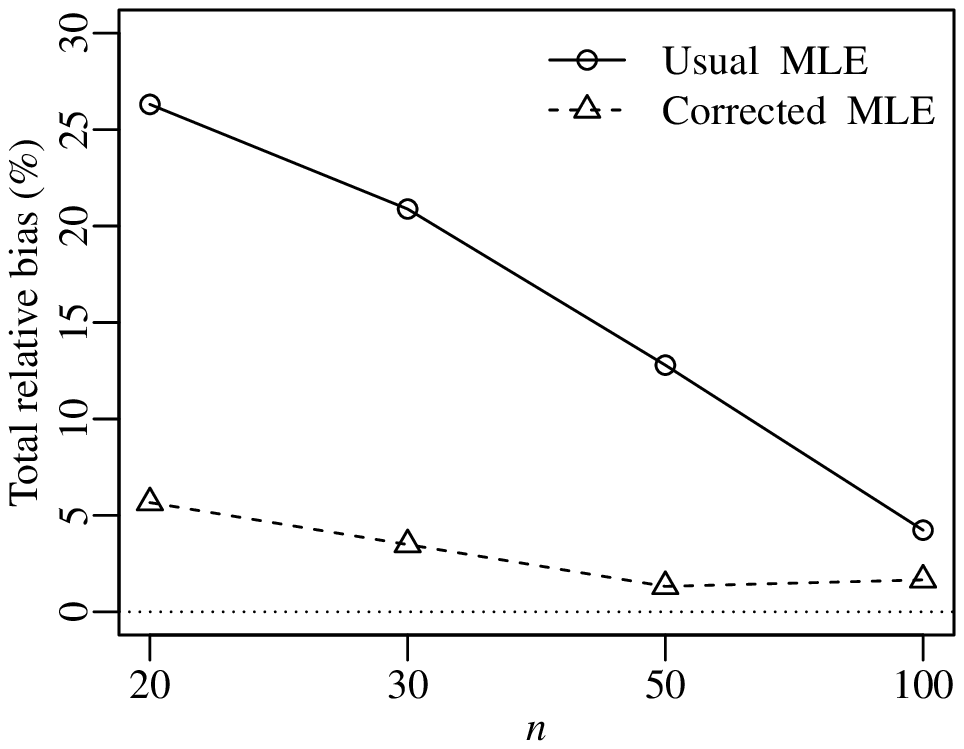}}
\subfigure[$\beta$MA$(1)$ model]
{\label{f:viesbma}\includegraphics[width=0.4\textwidth] {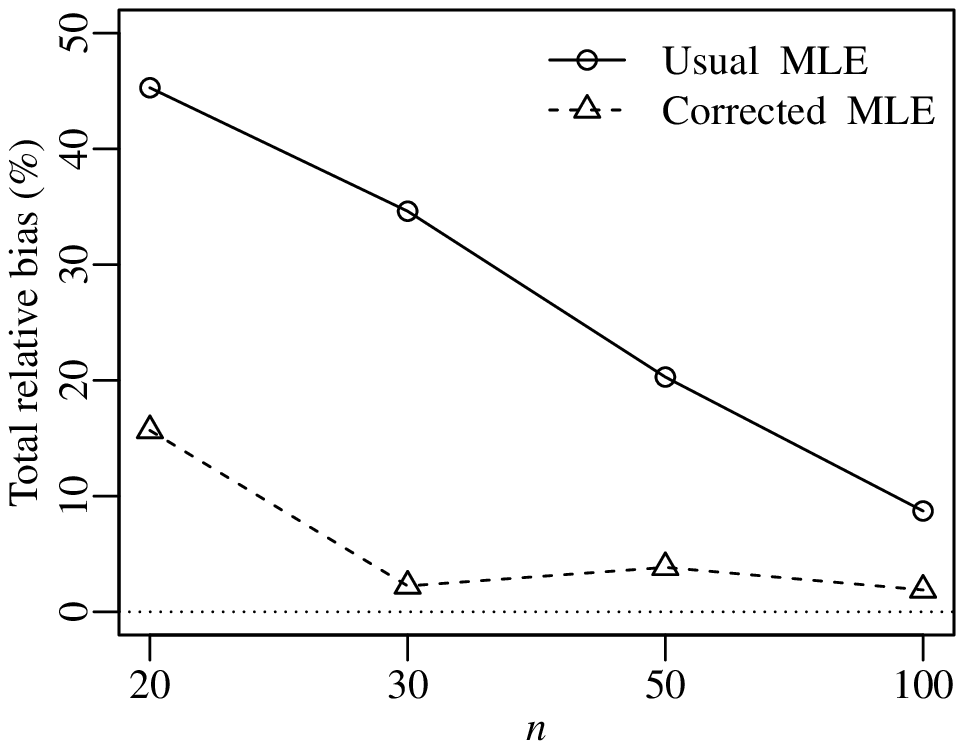}}\\
\subfigure[\barma$(1,1)$ model]
{\label{f:viesbarma1}\includegraphics[width=0.4\textwidth]{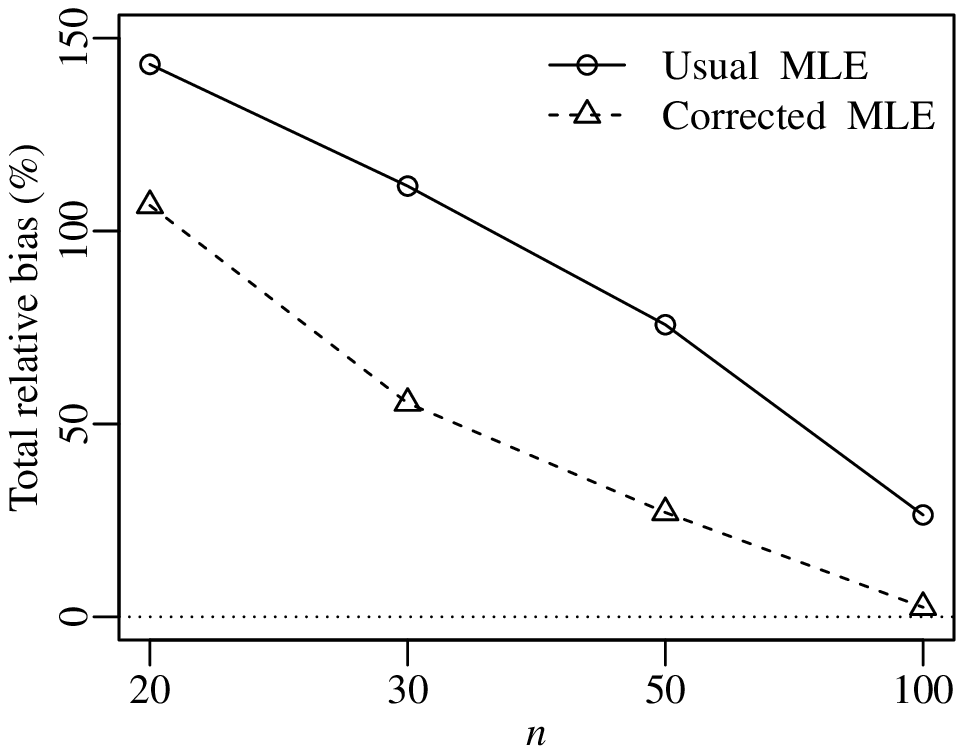}}
\caption{
Total relative bias of the estimator in several scenarios.
}\label{f:figvies} 
\end{center} 
\end{figure}

For the evaluation of interval estimation we calculate the coverage rates (CR) of each of the CI considered, 
with a significance of  $5\%$.
In each Monte Carlo replication we calculated the confidence interval and checked if the CI contains the true parameter or not.
The coverage rate is given by the percentage of replications in which the CI contained the parameter. It is desirable that the CR value gets closer to the
nominal coverage level $0.95$. Here are presented only graphical summaries of the numerical results, considering the average of coverage rates (ACR). Such ACR are calculated from the average of the coverage rates of all parameters of each CI. Thus, we can verify the general behavior of confidence intervals, but not on each parameter independently. In this graphical analysis we lose some information, but we gain in interpretability. 

Figure~\ref{f:figtaxa} shows graphically different ACR for the uncorrected and corrected estimators. In these results the lower distortion of the coverage rate of corrected CI becomes apparent when compared with the usual asymptotic CI.
Only the $\beta$AR$(1)$ model has reasonable results for the standard confidence interval. These results are in agreement with what is discussed in~\cite{Inoue2002}, where for models composed only for autoregressive parameter, the confidence interval shows good results. In other models the standard confidence intervals showed poor results, suggesting the necessity of corrections. The inferential difficulties on the parameters of the moving averages part are also verified in the classic article of \cite{Ansley1980}, considering ARMA models. In all models, the bootstrap corrections proved better than the usual interval estimators.
It can be concluded that the best corrected confidence interval is the $CI_{boot}$, once it presented values of coverage rates closer to 0.95 for all parameters in all sample sizes. 
It should be also noted that among the bootstrap confidence intervals, the $CI_u$ showed the greatest distortion in relation to the cover rates, although it presented less distortion than asymptotic intervals.

\begin{figure}[t] 
\begin{center} 
\subfigure[$\beta$AR$(1)$ model]
{\label{f:taxabar}\includegraphics[width=0.4\textwidth]{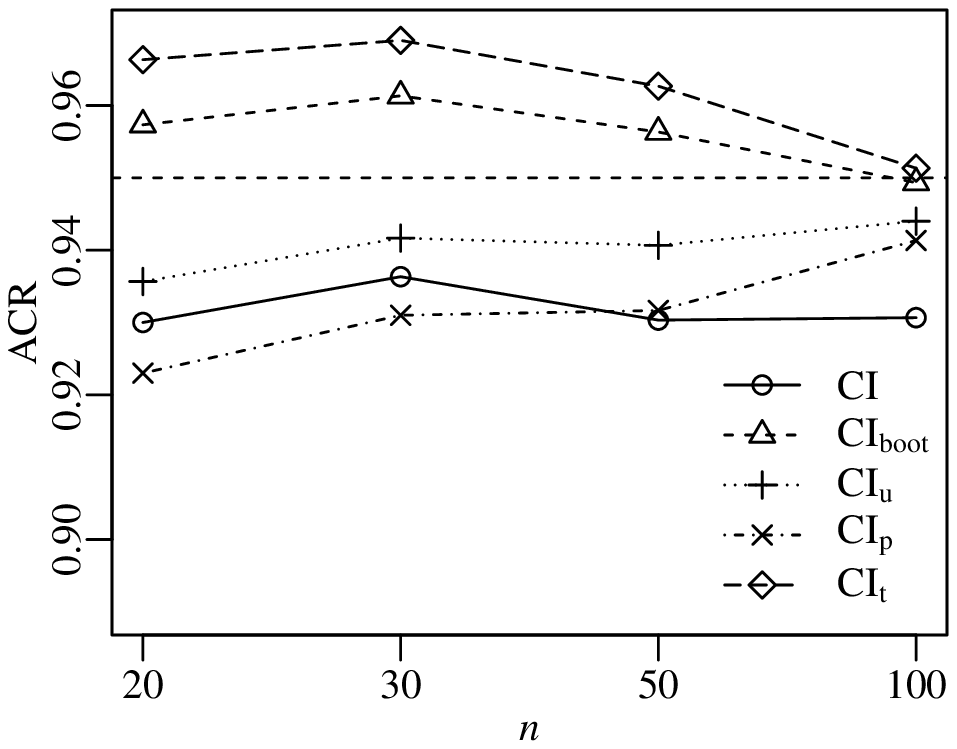}}
\subfigure[$\beta$MA$(1)$ model]
{\label{f:taxabma}\includegraphics[width=0.4\textwidth] {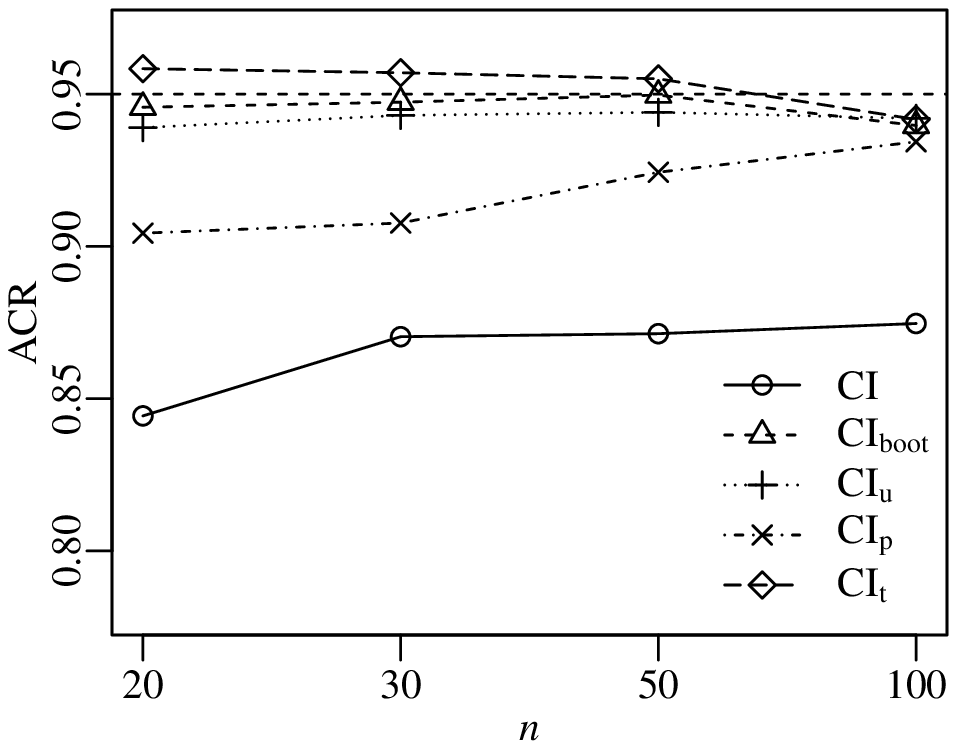}}\\
\subfigure[\barma$(1,1)$ model]
{\label{f:taxabarma1}\includegraphics[width=0.4\textwidth]{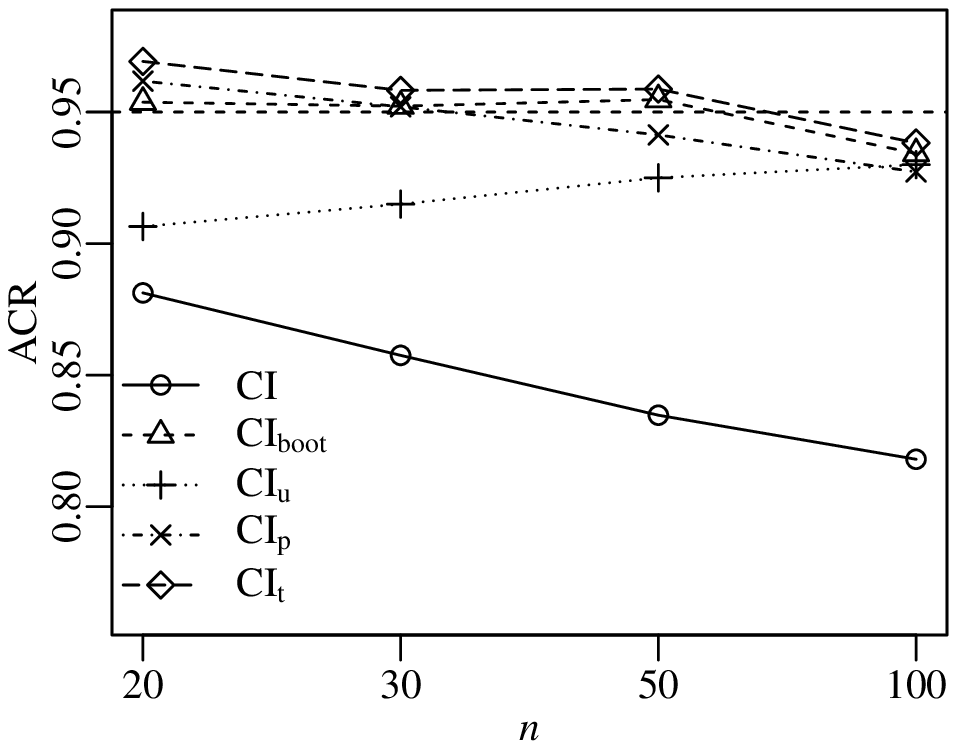}}
\caption{
Average coverage rates for several confidence intervals in different scenarios.
}\label{f:figtaxa} 
\end{center} 
\end{figure}

\section{Application}\label{s:aplicacao} 

This section presents an application to real data of the \barma~model 
with the corrected estimators proposed in Section~\ref{s:corr}. We considered comparisons among the forecast of the corrected \barma, the uncorrected \barma, and the classical ARMA model \citep{Box2008}. The data used refer to the percentage of energy stored in southern Brazil, in the period of January $2009$ to January $2015$, totaling $73$ observations \citep{ONS2014}. 
The study of the amount of stored energy becomes an important aid to managers in the water resources and electricity areas.
Such forecasts are useful to predict problems of lack or accumulation of energy,
and could avoid waste. Thus, one can also control the expenses and, consequently, prices \citep{Hong2014}. 

An implementation in {\tt R} language~\citep{R2012}  to fit \barma~model 
with the proposed corrected MLE and diagnostic tools is available at 
\url{http://www.ufsm.br/bayer/boot-barma.zip}. 
The file contains computer codes and also the dataset used in this empirical application.

{The series mean equals $0.6955$, its standard deviation being equal to $0.2078$ and the coefficient of variation being equal to $0.0432$. The maximal and minimal values are $0.9789$ and $0.3461$, respectively.} The time series can be observed graphically in Figure~\ref{f:serie}, while the histogram of the data is presented in Figure~\ref{f:hist}. The histogram shows an asymmetric behavior of the data distribution, reinforcing that the choice of a model that assumes normality, such as ARMA model, would be inappropriate. The sampling autocorrelation function (ACF) and the sampling partial autocorrelation function (PACF) are shown in Figures \ref{f:facdados} and \ref{f:facpdados}, respectively. 

\begin{figure}[t] 
\begin{center} 
\subfigure[Original series]
{\label{f:serie}\includegraphics[width=0.4\textwidth]{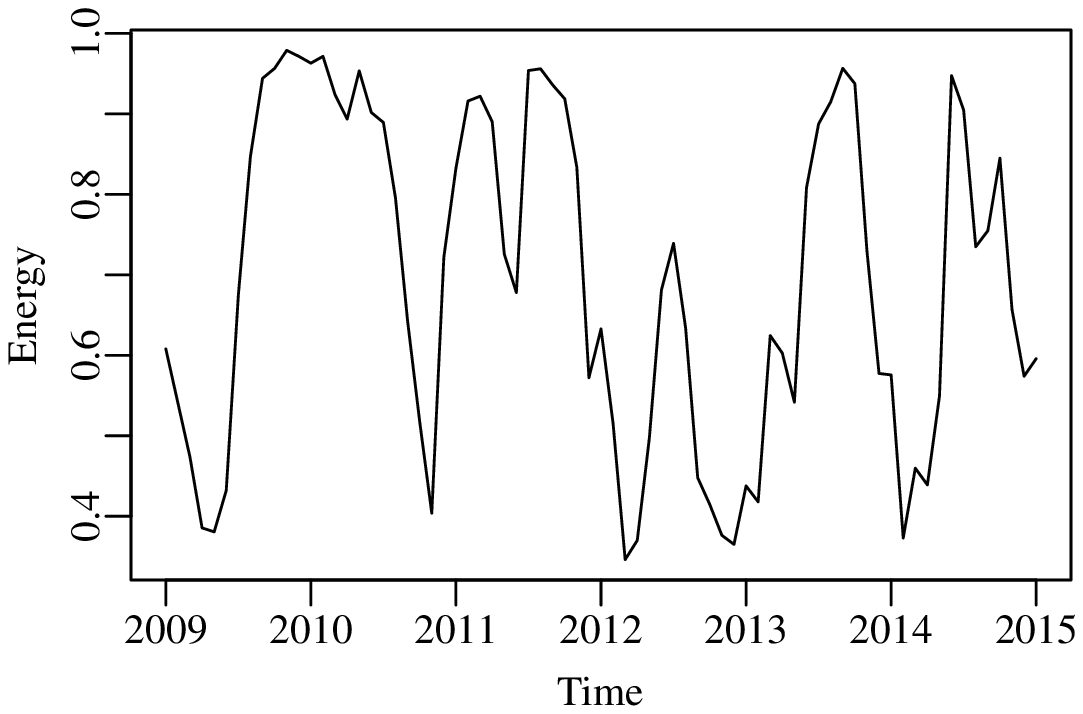}} 
\subfigure[Histogram]
{\label{f:hist}\includegraphics[width=0.4\textwidth]{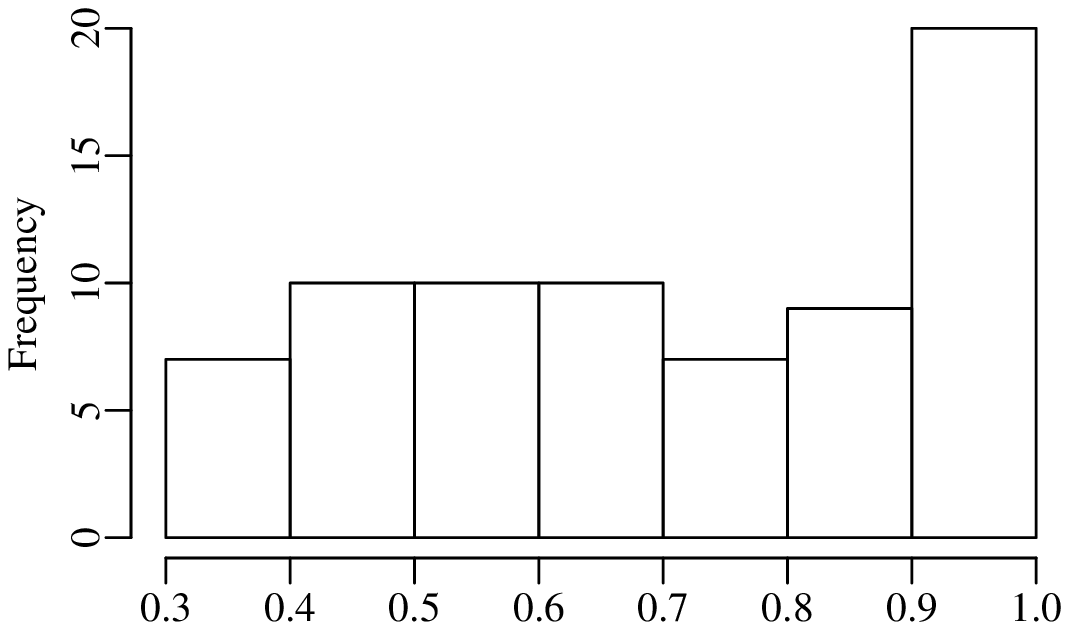}} 
\subfigure[Sampling ACF]
{\label{f:facdados}\includegraphics[width=0.4\textwidth] {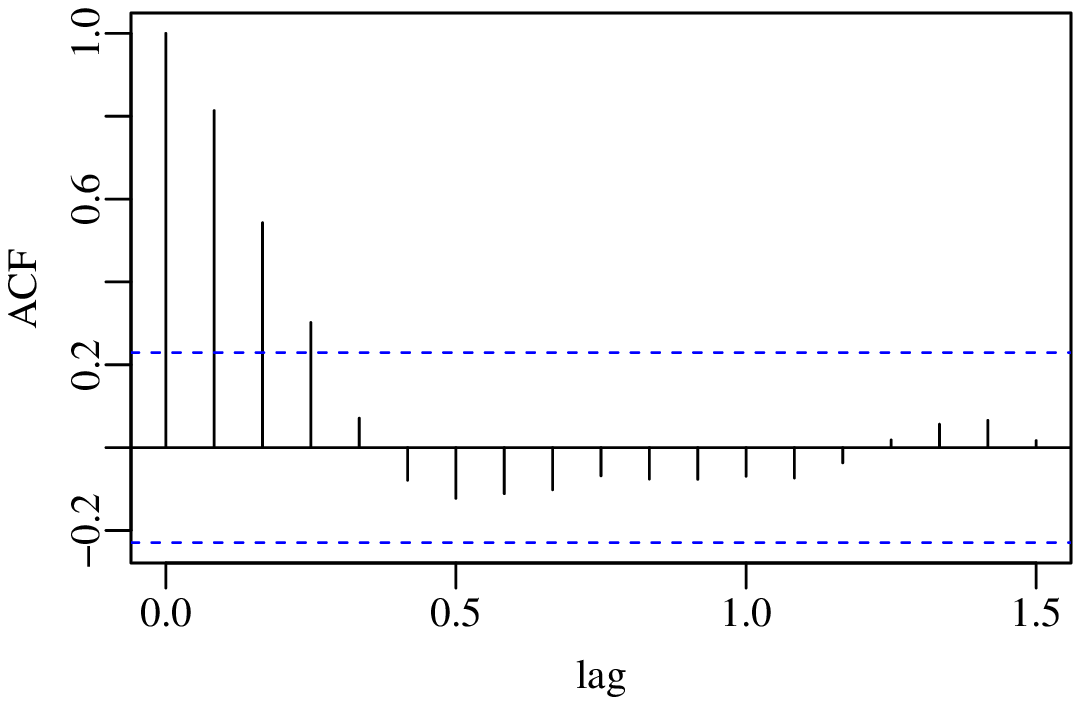}}
\subfigure[Sampling PACF]
{\label{f:facpdados}\includegraphics[width=0.4\textwidth] {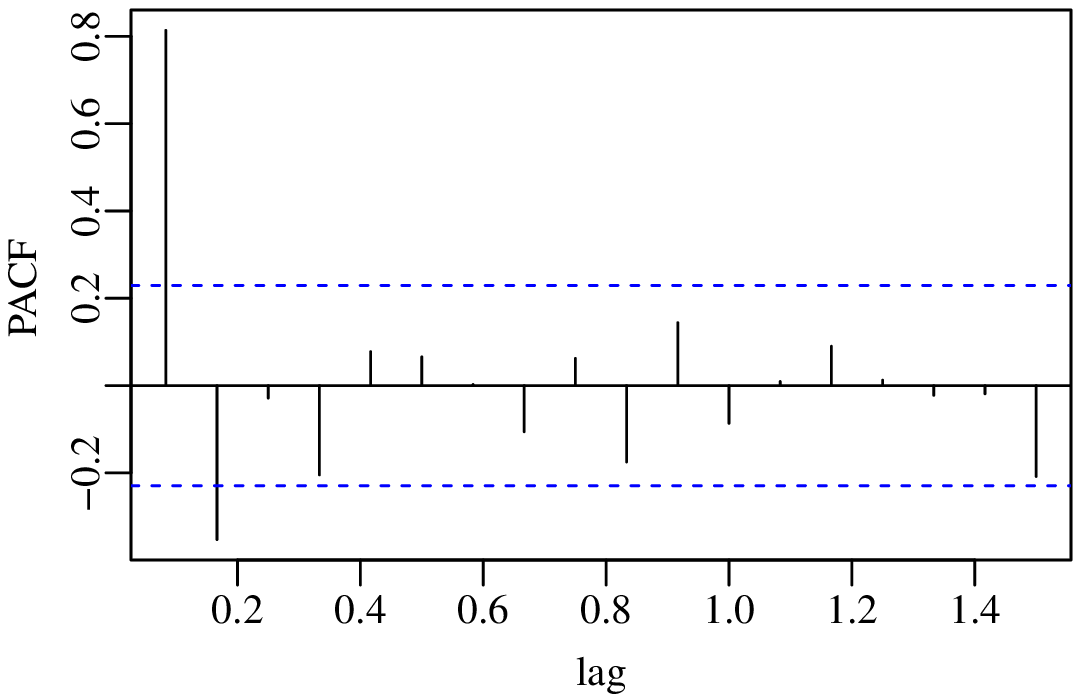}}
\caption{0.346
Line chart, histogram and correlograms of the time series of percentage of stored energy.
}\label{f:dados} 
\end{center} 
\end{figure} 

For the model selection it was considered a computationally exhaustive search with the objective of minimizing the Akaike information criterion (AIC)~\citep{Akaike}. The search space was restricted to models with orders less than or equal to $6$, that is, we considered $0 \leq p \leq 6$ and $0 \leq q \leq 6.$ The \barma$(1,1)$ model presented
the lowest AIC value. 
{The logit link function was considered.}

Table~\ref{T:ajuste1} presents the fit of the selected model, considering the point and interval estimates (with $\alpha=5\%$) corrected and uncorrected. The corrected CI considered was the $CI_{boot}$, because it was the CI that presented best 
numerical results in simulations. 
Zero does not belong to any 95\% level confidence interval, and hence all the parameters are statistically significant at the 5\% nominal level. 
Based on the numerical results, which showed considerable biases of MLE and the effectiveness of the bootstrap correction, it is suggested that the bias-corrected estimates result in a model closer to the unknown population model and, therefore, in a better fitted model. 

\begin{table}[t] 
\caption{\barma$(1,1)$ model adjusted for the stored energy data, considering corrected and uncorrected estimators.} 
\label{T:ajuste1} 
\tablesize
\begin{center}
\begin{tabular}{rrrrrr} 
 \hline
{Adjustments}&&${\alpha}$ & ${\varphi}_1$ &  ${\theta}_1$ & ${\phi}$\\ 
\hline
Usual MLE && $0.3222$ & $0.5746$ &  $2.4871$  & $13.3225$\\
Corrected MLE &&  $0.2713$ & $0.6503$ & $2.4206$ & $12.3485$ \\
$CI$ &$L$ & $0.1483$ & $0.4844$  & $1.4205$   & $9.0288$ \\
 & $U$   & $0.4961$ &  $0.6647$  & $3.5538$ & $17.6162$    \\
$CI_{boot}$ &$L$ & $0.0090$ & $0.3693$ & $0.6514$ & $0.7387$ \\
& $U$  & $0.6354$ & $0.7798$ & $4.3229$  & $25.9062$ \\

\hline
 
\end{tabular}	
\end{center}	
\end{table}

Figure~\ref{f:res} shows some useful charts of the fitted \barma$(1,1)$ model with corrected estimators and diagnostic analysis based on the standardized residuals. 
The investigation of graphical analysis shows acceptable fit, since the residual correlograms don't show autocorrelation significantly different from zero, the behavior of the standardized residuals is random around zero and within the interval of $-3$ to $3$ and the predicted values are similar to the original data.

\begin{figure}[t] 
\begin{center} 
\subfigure[Predicted values]
{\label{f:adjusted}\includegraphics[width=0.4\textwidth] {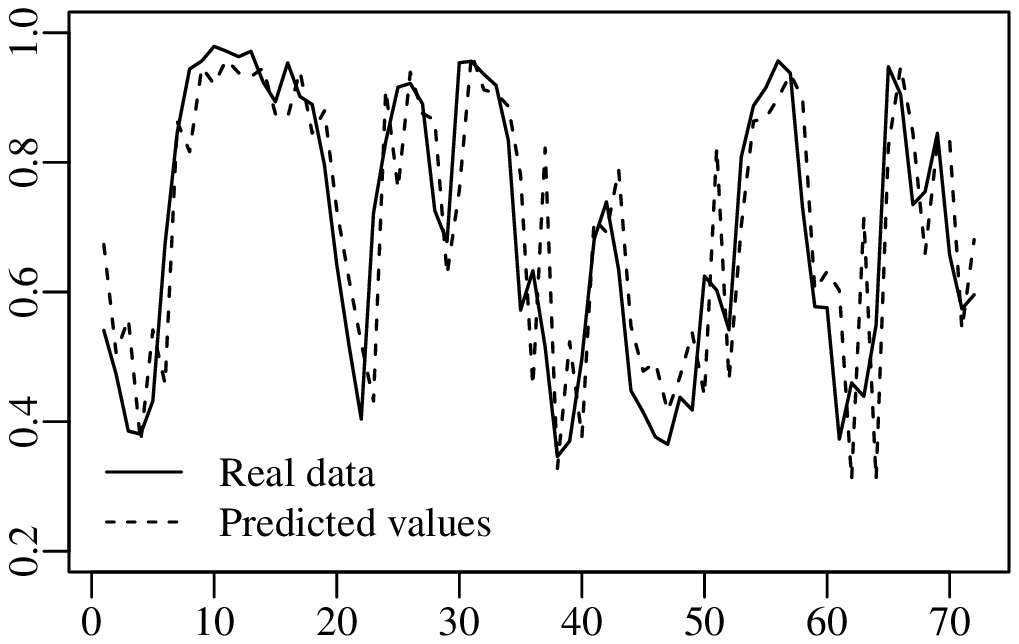}}
\subfigure[Standardized residual]
{\label{f:rescor}\includegraphics[width=0.4\textwidth] {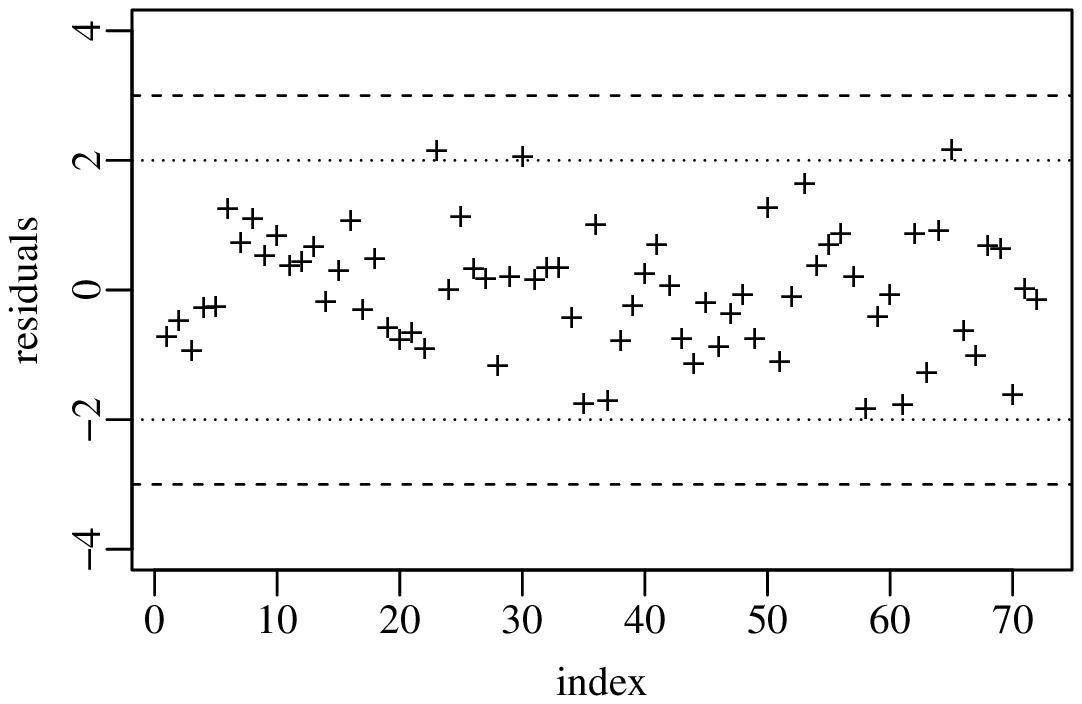}}\\
\subfigure[Residual ACF]
{\label{f:fac}\includegraphics[width=0.4\textwidth]{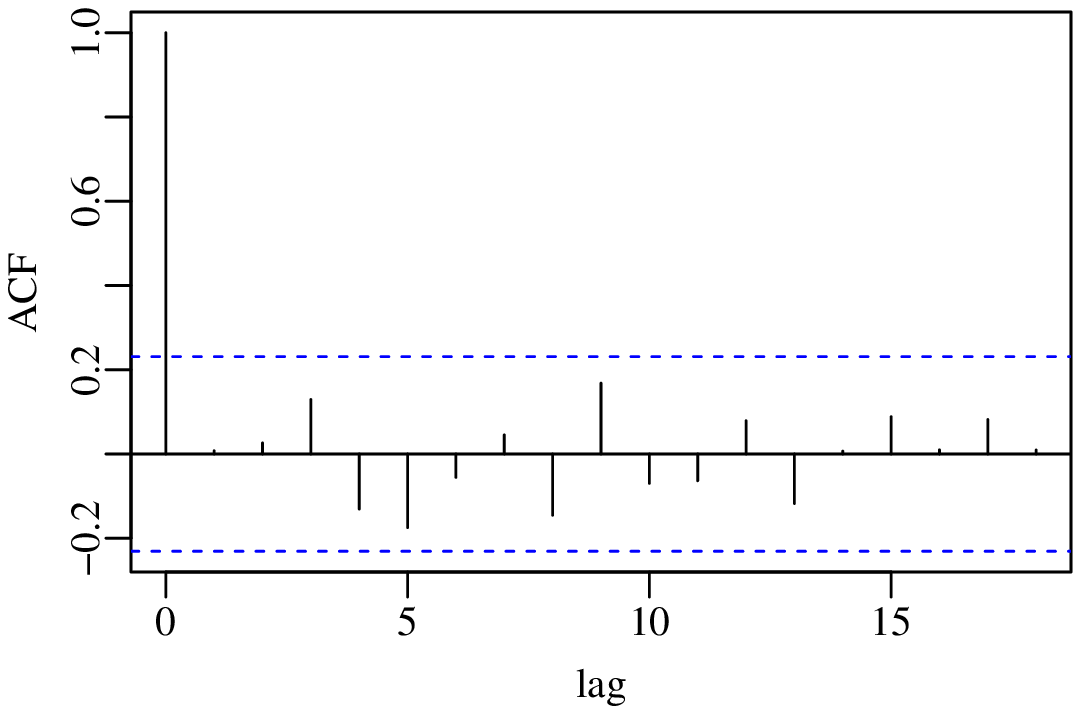}}
\subfigure[Residual PACF]
{\label{f:facp}\includegraphics[width=0.4\textwidth] {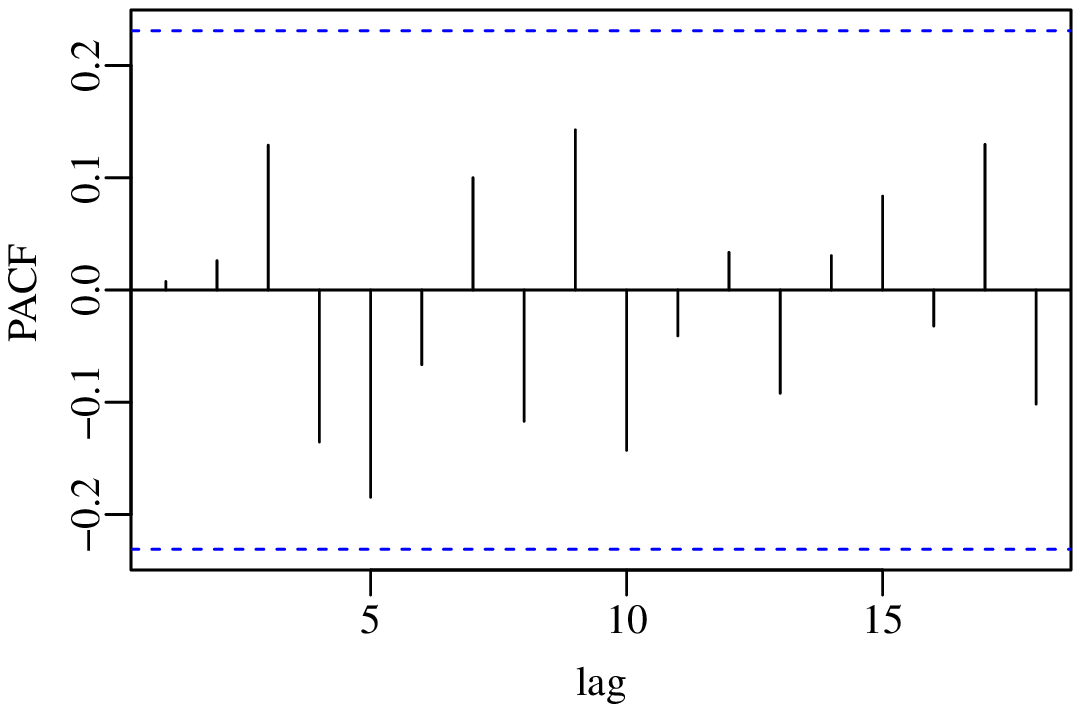}}
\caption{Charts of the fitted \barma$(1,1)$ model with bias-corrected estimators.}\label{f:res}
\end{center} 
\end{figure} 

Aiming to empirically verify the importance of considering bias-corrected estimators, we carried out comparisons of the models in relation to actual and predicted values. These comparisons were performed by evaluating the mean square error (MSE), the mean absolute percentage error (MAPE) and the mean absolute scaled error (MASE)~\citep{Hyndman2006}. 
The MSE and the MAPE are traditional quality measures of prediction
and MASE is indicated for values close to zero~\citep{Hyndman2006}. These measures evaluate the difference between the actual value and the predicted value and are defined respectively by:  
\begin{align*}
{\rm MSE} & = \frac{1}{H}\sum \limits _{h=1}^H (y_h-\widehat{\mu}_h)^2, \\
{\rm MAPE} &= \frac{1}{H}\sum \limits _{h=1}^H \frac{\vert y_h-\widehat{\mu}_h\vert}{\vert y_h \vert}, \, \text{ and }\\
{\rm MASE} &=  \frac{1}{H}\sum \limits _{h=1}^H  \left( \frac{\vert y_h-\widehat{\mu}_h\vert}{\frac{1}{h-1}\sum \limits _{h=2}^H \vert y_h-y_{h-1}\vert}\right) \text,
\end{align*}
where $y_h$ are the observed data, $\widehat{\mu}_h$ are the predicted values and $h=1,\ldots,H$, with $H=6$. Such measures were adjusted within and outside the interval of observations used for estimation. The calculation with data outside of the observational period was carried out considering the $H$ last ocurrences reserved from the original data. Table~\ref{T:ajuste18} presents the actual and predicted values for the referred months for corrected and uncorrected \barma~models. Indeed, we considered the ARMA(1,1) model for a simple predictive capability comparison. It can be seen that the corrected \barma~model had predicted values closer to the observed data than the other two models. It is also noticeable that MSE, MAPE and MASE measures are smaller when the \barma~with corrected estimators are considered, highlighting the better fit. 

\begin{table}[t]	
\caption{Actual and predicted values for the different models.} 
\label{T:ajuste18} 
\tablesize
\begin{center} 
\begin{tabular}{rcccc} 
\hline
	&	 Observed  	&	 Corrected 	&	 Uncorrected 	&	 Classical \\
Months 	&	 data 	&	 $\beta$ARMA  	&	 $\beta$ARMA 	&	 ARMA \\    
\hline								
February  (2015) 	& $0.5111$ 	& $0.6166$ 	& $0.6185$ 	&	 $0.6228$ \\			
March  (2015) 	&	$0.3930$ 	&	 $0.6411$ 	&	 $0.6456$ 	& $0.6445$ \\						
April (2015) 	& $0.3416$ 	&	 $0.6567$ 	&	  $0.6608$ 	&	 $0.6593$ \\					
May (2015) 	& $0.3807$ 	&	 $0.6667$ 	&	   $0.6694$ 	&	 $0.6693$ \\								
June  (2015) 	&	$0.6367$ 	&	 $0.6730$ 	&	   $0.6742$ 	&	 $0.6762$ \\
								
July (2015) 	& $0.9676$ 	&	 $0.6772$ 	&	   $0.6770$ 	&	 $0.6808$ \\
\hline								
&	 MSE 	&	 $0.0818$ 	&	 $0.0851$ 	&	 $0.0869$\\
&	 MAPE 	&	 $0.4781$ 	&	 $0.4841$ 	&	 $0.4842$\\
&	 MASE 	&	 $1.3424$	&	 $1.3575$ 	&	 $1.3575$\\
\hline
\end{tabular}	
\end{center}	
\end{table}

Still, within the period analyzed, with 73 observations, the model with corrected estimators demonstrated more accurate results. In this case, the values of MSE, MAPE and MASE were, respectively, 
$0.0000$, $0.1427$ and $0.9290$ for the \barma~with corrected estimators, 
$0.0002$, $0.1473$ and $0.9559$ for the uncorrected \barma~model,  
and $0.0001$, $0.1450$ and $0.9556$ for the ARMA model.
In general, there are improvements in fit and forecasts for the model with corrected estimators. This can also be seen in works that consider another class of time series models, as in~\cite{Kim2003} and \cite{Kim2012}. The non corrected models demonstrated inferior quality measurements for both inside and outside used intervals fitting modeling. Therefore, the effectiveness of the proposed correction was verified.

\section{Conclusions} \label{s:con} 

The ARIMA modeling is commonly used for modeling and forecasting variables over time. However, these models become inappropriate when it is not reasonable to assume normality for the variable of interest  $y$, specially when $y$ belongs in the continuous interval ($0,1$). For such situations, there is the \barma~model, assuming beta distribution for the variable of interest. The parameters of this model are estimated via maximization of the conditional log-likelihood function. These inferences have good results when the sample size is large, but can lead to inferential distortions in small samples. In the present paper were proposed bootstrap corrections for point and interval small sample inferences in the \barma~model. The evaluation of the corrections considered was developed through Monte Carlo simulations and in an empirical application. The numerical results indicate that the corrected estimates are less biased, decreasing or nullifying the distortion problem in small samples. Moreover, we verified improvement in values of corrected confidence intervals. 
An application to real data was considered, observing predicted values closer to the actual values observed in the model with corrected estimators. Finally, we suggest the use of bootstrap corrections for small sample inferences in \barma~model.

\section*{Acknowledgements}

We gratefully
acknowledge partial financial support from 
Funda\c{c}\~ao de Amparo \`a Pesquisa do Estado do Rio Grande do Sul (FAPERGS), Conselho Nacional de Desenvolvimento Cient\'ifico e Tecnol\'ogico (CNPq), and Coordena\c{c}\~ao de Aperfei\c{c}oamento de Pessoal de N\'ivel Superior (CAPES), Brazil.

\singlespacing

\bibliographystyle{arxiv}

\bibliography{betareg}

\appendix 

\section*{Appendix}

In this appendix, we present the scenarios of simulations that were not discussed in the text. 
The numerical results of point and interval estimates of the $\beta$AR$(1)$, $\beta$MA$(1)$ and \barma$(1,1)$ models, with $\phi=120$, are presented.

\begin{table}[h]
\caption {Results of the Monte Carlo simulation for point estimation in $\beta$AR$(1)$ model.}
\tablesizea
\label{T:bar1201}
\begin{center}
\begin{tabular}{rrrrrrr}
\hline
Measures 	&	$\widehat{\alpha}$ 	&	$\overline{\alpha}$  	&	 $\widehat{\varphi}_{1}$ 	&	 $\overline{\varphi}_{1}$  	&	$\widehat{\phi}$ 	&	$\overline{\phi}$		\\                      
\hline                 
\multicolumn{7}{c}{$n=20$}\\ 
\hline	
Mean  & $0.991$  &    $0.999$ & $-0.484$  &  $-0.497$ & $132.552$ &  $117.489$\\
Bias  & $-0.009$  &   $-0.001$ &  $0.016$  &   $0.003$  & $12.552$ &   $-2.511$\\
RB    & $-0.937$   &  $-0.052$ & $-3.293$  &  $-0.555$ &  $10.460$  &  $-2.093$\\
SE   &   $0.135$   &   $0.151$ &  $0.191$  &   $0.215$ &  $51.500$  &  $45.661$\\
MSE   &  $0.018$   &   $0.023$ &  $0.037$   &  $0.046$ &$2809.835$ & $2091.242$\\
\hline
\multicolumn{7}{c}{$n=30$}\\                           
\hline  
Mean  & $0.996$  &    $1.004$ & $-0.496$  &  $-0.508$ & $126.536$ &  $118.625$\\
Bias &  $-0.004$  &    $0.004$ &  $0.004$  &  $-0.008$  &  $6.536$ &   $-1.375$\\
RB    & $-0.421$   &   $0.377$ & $-0.816$  &   $1.596$  &  $5.447$  &  $-1.146$\\
SE  &    $0.110$  &    $0.119$ &  $0.153$  &   $0.167$  & $36.908$ &   $34.315$\\
MSE   &  $0.012$  &    $0.014$ &  $0.023$   &  $0.028$ &$1404.938$ & $1179.397$\\
\hline
\multicolumn{7}{c}{$n=50$}\\                           
\hline
Mean  & $0.988$  &    $0.993$ & $-0.484$  &  $-0.491$ &$123.221$ &  $119.763$\\
Bias &  $-0.012$   &  $-0.007$ &  $0.016$  &   $0.009$  & $3.221$  &  $-0.237$\\
RB   &  $-1.204$  &   $-0.723$ & $-3.285$  &  $-1.826$ &  $2.684$ &   $-0.198$\\
SE  &    $0.086$   &   $0.090$  & $0.123$  &   $0.130$ & $27.686$ &   $26.585$\\
MSE  &   $0.008$   &   $0.008$ &  $0.015 $ &   $0.017$ &$776.885$ &  $706.818$\\
\hline
\multicolumn{7}{c}{$n=100$}\\                           
\hline
Mean &  $0.994$   &   $0.997$ & $-0.491$  &  $-0.495$ &$119.835$ &  $119.197$\\
Bias &  $-0.006$  &   $-0.003$ &  $0.009$   &  $0.005$ & $-0.165$ &   $-0.803$ \\
RB  &   $-0.600$  &   $-0.305$ & $-1.862$ &   $-1.004$ & $-0.137$ &   $-0.669$\\
SE   &   $0.059$  &    $0.060$ &  $0.082$  &   $0.084$ & $18.211$  &  $17.908$\\
MSE  &  $ 0.003$  &    $0.004$ &  $0.007$  &   $0.007$ &$331.653$ &  $321.338$\\
\hline
\end{tabular}			
\end{center}			
\end{table}
\begin{table}	
\caption {Results of the Monte Carlo simulation for point estimation in $\beta$MA$(1)$ model.}
\tablesizea
\label{T:bma1201}
\begin{center}
\begin{tabular}{rrrrrrr}
\hline
Measures 	&	$\widehat{\alpha}$ 	&	$\overline{\alpha}$  	&	 $\widehat{\theta}_{1}$ 	&	 $\overline{\theta}_{1}$  	&	$\widehat{\phi}$ 	&	$\overline{\phi}$		\\                      
\hline                 
\multicolumn{7}{c}{$n=20$}\\ 
\hline	
Mean & $-0.997$  &   $-0.998$ &   $0.647$  &     $1.017$ & $132.965$ &  $125.316$\\
Bias &   $0.003$   &   $0.002$  & $-0.353$   &    $0.017$ &  $12.965$ &    $5.316$\\
RB   &  $-0.317$   &  $-0.181$ & $-35.285$  &     $1.708$ &  $10.804$ &    $4.430$\\
SE  &    $0.056$   &   $0.057$  &  $1.431$    &   $1.554$  & $55.179$  &  $52.205$\\
MSE  &   $0.003$   &   $0.003$  &  $2.172$   &    $2.417$ &$3212.765$ & $2753.588$\\
\hline
\multicolumn{7}{c}{$n=30$}\\                           
\hline  
Mean  &$-0.998$  &   $-1.000$  &  $0.819$  &     $1.036$ & $124.234$ &  $121.463$\\
Bias  &  $0.002$   &   $0.000$  & $-0.181$  &     $0.036$  &  $4.234$ &    $1.463$\\
RB   &  $-0.154$    & $-0.042$ & $-18.099$  &     $3.567$ &   $3.528$ &    $1.219$\\
SE  &    $0.045$   &   $0.045$  &  $1.043$  &     $1.068$ &  $36.102$ &   $34.956$\\
MSE   &  $0.002$    &  $0.002$   & $1.121$   &    $1.141$ &$1321.289$ & $1224.069$\\
\hline
\multicolumn{7}{c}{$n=50$}\\                           
\hline
Mean & $-1.001$  &   $-1.002$ &   $0.897$  &     $0.998$ &$120.438$ &  $121.343$\\
Bias &  $-0.001$  &   $-0.002$ &  $-0.103$  &    $-0.002$ &  $0.438$ &    $1.343$\\
RB     & $0.099$   &   $0.193$ & $-10.320$  &    $-0.240$ &  $0.365$ &    $1.119$\\
SE  &    $0.036$   &   $0.036$  &  $0.798$   &    $0.787$  &$26.607$  &  $26.554$\\
MSE    & $0.001$  &    $0.001$  &  $0.648$   &    $0.620$ &$708.098$  & $706.913$\\
\hline
\multicolumn{7}{c}{$n=100$}\\                          
\hline
Mean & $-0.998$  &   $-0.999$  &  $0.954$   &    $1.009$ &$117.313$ &  $119.987$\\
Bias   & $0.002$  &    $0.001$ &  $-0.046$   &    $0.009$ & $-2.687$  &  $-0.013$\\
RB  &   $-0.155$   &  $-0.082$ &  $-4.599$  &     $0.910$ & $-2.239$  &  $-0.011$\\
SE  &    $0.025$   &   $0.025$   & $0.515$  &     $0.517$ & $17.519$  &  $17.861$\\
MSE  &   $0.001$   &   $0.001$   & $0.267$   &    $0.267$ &$314.152 $ & $319.032$\\
\hline
\end{tabular}			
\end{center}			
\end{table}
\begin{table}	
\caption {Results of the Monte Carlo simulation for point estimation in \barma$(1,1)$ model.}
\tablesizea
\label{T:barma1}
\begin{center}
\begin{tabular}{rrrrrrrrr}
\hline
Measures 	&	$\widehat{\alpha}$ 	&	$\overline{\alpha}$  	&	$\widehat{\varphi}_{1}$ 	&	 $\overline{\varphi}_{1}$ &$\widehat{\theta}_{1}$ 	&	 $\overline{\theta}_{1}$  	&	$\widehat{\phi}$ 	&	$\overline{\phi}$		\\                      
\hline                 
\multicolumn{9}{c}{$n=20$}\\ 
\hline	
Mean  & $1.298$   &   $1.161$ &  $0.350$  &   $0.423$  & $-1.183$ &     $-1.367$ & $135.242$ &$135.016$\\
Bias &   $0.298$   &   $0.161$ & $-0.150$  &  $-0.077$  &  $0.317$ &      $0.133$ &  $15.242$ & $15.016$\\
RB    & $29.766$   &  $16.071$ &$-30.048$  & $-15.393$ & $-21.131$  &    $-8.850$ &  $12.702$ &$12.514$\\
SE &  $0.678$   &   $0.851$  & $0.346$   &  $0.442$  &  $4.571$ &      $6.356$ &  $52.549$& $60.639$\\
MSE  &   $0.549$   &   $0.750$  & $0.143$   &  $0.201$  & $20.998$  &    $40.410$ &$2993.746$ &$3902.615$\\
\hline
\multicolumn{9}{c}{$n=20$}\\ 
\hline	
Mean  &$1.278$   &   $1.145$ &  $0.363$  &   $0.431$ &  $-0.721$  &    $-1.048$ & $130.596$ & $129.497$\\
Bias &    $0.278$  &   $ 0.145$ & $-0.137$  &  $-0.069$  &  $0.779$  &    $ 0.452$  & $10.596$ & $9.497$\\
RB    & $27.815$  &   $14.541$ &$-27.300$ &  $-13.784$  &$-51.964$ &   $ -30.111$  &  $8.830$ &$7.914$\\
SE &  $0.615 $  &   $0.764$ &  $0.307$  &   $0.383$  &  $3.779$  &    $ 4.972$ & $ 38.917$ & $40.668$\\
MSE  &    $0.455$  &    $0.604$ &  $0.113$  &   $0.151$  & $14.888$  &   $ 24.925$ & $1626.789$ & $1744.050$\\
\hline
\multicolumn{9}{c}{$n=50$}\\                           
\hline
Mean &  $1.227$  &    $1.095$  & $0.389$  &   $0.454$  & $-0.703$  &    $-1.129$ &$124.484$ &$123.994$\\
Bias   & $0.227$  &    $0.095$ & $-0.111$  &  $-0.046$  &  $0.797$  &     $0.371$ &  $4.484$ &$3.994$\\
RB    & $22.701$   &   $9.453$ &$-22.292$  &  $-9.159$ & $-53.110$  &   $-24.729$ &  $3.737$ &$3.328$\\
SE  &   $0.509$   &   $0.594$  & $0.252$   &  $0.295$  &  $2.825$   &    $3.357$ & $27.700$ &$28.962$\\
MSE    & $0.310$   &   $0.362$  & $0.076$  &   $0.089$  &  $8.613$  &    $11.404$ &$787.386$& $854.771$\\
\hline   
\multicolumn{9}{c}{$n=100$}\\                           
\hline
Mean  & $1.161$   &   $1.047$ &  $0.421$  &   $0.477$  & $-0.856$  &    $-1.295$ &$120.976$ &$121.274$\\
Bias  & $0.161$   &   $0.047$ & $-0.079$  &  $-0.023$  & $0.644$  &     $0.205$ &  $0.976$ &$1.274$\\
RB    & $16.141$  &    $4.689$ &$-15.833$  &  $-4.564$ & $-42.916$  &   $-13.688$ &  $0.813$ &$1.062$\\
SE   &   $0.395$  &    $0.450$ &  $0.194$  &   $0.221$  &  $2.077$   &    $2.384$ & $17.826$ &$18.069$\\
MSE    & $0.182$  &    $0.204$ &  $0.044$  &   $0.049$  &  $4.727$  &     $5.726$ &$318.723$ &$328.099$\\
\hline
\end{tabular}			
\end{center}		
\end{table}
\begin{table}	
\caption {Average coverage rates for several confidence intervals in $\beta$AR$(1)$ model.}
\tablesizea
\label{T:bar1202}
\begin{center}
\begin{tabular}{rrrr}
\hline
Intervals 	&	${\alpha}$  &  ${\varphi}_{1}$ & ${\phi}$ \\
\hline 
\multicolumn{4}{c}{$n=20$}\\ 
\hline
$IC$  &            $0.947$ &$0.951$ &$0.947$\\
$CI_{boot} $  &      $ 0.931$ &$0.938$ &$0.979$\\
$CI_{u}$ &  $0.912$ &$0.897$ &$0.967$\\
$CI_{p}$ & $0.964$ &$0.971$ &$0.936$\\
$CI_{t}$   &     $0.951$ &$0.952$ &$0.985$\\	      
\hline 
\multicolumn{4}{c}{$n=30$}\\
\hline                
$IC$  &            $0.948$ &$0.955$ &$0.937$\\
$CI_{boot} $  &     $0.942$ &$0.948$ &$0.969$\\
$CI_{u}$ & $0.924$ &$0.925$ &$0.957$\\
$CI_{p}$ & $0.963$ &$0.968$ &$0.944$\\
$CI_{t}$   &         $0.953$ &$0.953$ &$0.975$\\ 
\hline
\multicolumn{4}{c}{$n=50$}\\                           
\hline
$IC$  &           $0.945$ &$0.946$ &$0.936$\\
$CI_{boot} $  &      $0.942$ &$0.938$ &$0.956$\\
$CI_{u}$ &  $0.929$ &$0.920$ &$0.949$\\
$CI_{p}$ &$0.954$ &$0.956$ &$0.940$\\
$CI_{t}$     &    $0.951$ &$0.947$ &$0.958$\\
\hline
\multicolumn{4}{c}{$n=100$}\\                           
\hline
$IC$  &           $0.958$ &$0.960$ &$0.932$\\
$CI_{boot} $  &      $0.955$ &$0.958$ &$0.941$\\
$CI_{u}$ &  $0.953$&$ 0.953$ &$0.943$ \\
$CI_{p}$ & $0.962$ &$0.967$ &$0.943$\\
$CI_{t}$   &         $0.958$ &$0.960$&$ 0.946$\\
\hline
\end{tabular}			
\end{center}			
\end{table}
\begin{table}	
\caption {Average coverage rates for several confidence intervals in $\beta$MA$(1)$ model.}
\tablesizea
\label{T:bma1202}
\begin{center}
\begin{tabular}{rrrr}
\hline
Intervals 	&	${\alpha}$  &  ${\varphi}_{1}$ & ${\phi}$ \\        
\hline 
\multicolumn{4}{c}{$n=20$}\\
\hline                
$IC $   &            $0.791$ &$ 0.872$ &$0.926$\\
$CI_{boot} $ &     $0.895$ & $0.951$ &$0.964$\\
$CI_{u}$ &  $0.896$ & $0.936$ &$0.947$\\
$CI_{p}$ &$0.900$ & $0.916$ &$0.924$\\
$CI_{t}$  &        $0.913$ &$ 0.963$& $0.972$\\
\hline
\multicolumn{4}{c}{$n=30$}\\                           
\hline
$IC $   &          $0.815$ & $0.873$ &$0.934$\\
$CI_{boot} $ &      $0.933$ & $0.946$ &$0.963$\\
$CI_{u}$ &  $0.932$ & $0.930$ &$0.957$\\
$CI_{p}$ & $0.935$ & $0.919$ &$0.944$\\
$CI_{t}$  &         $0.939$ & $0.957$ &$0.971$\\
\hline
\multicolumn{4}{c}{$n=50$}\\                           
\hline
$IC $   &       $0.800$ & $0.874$ &$0.909$\\
$CI_{boot} $ &    $0.925$ & $0.943$ &$0.932$\\
$CI_{u}$ &$0.927$ & $0.931$ &$0.941$\\
$CI_{p}$ &$0.928$ & $0.920$ &$0.932$\\
$CI_{t}$  &       $0.939$ & $0.949$ &$0.936$\\
\hline
\multicolumn{4}{c}{$n=100$}\\                           
\hline
$IC $   &     $0.808$ & $0.899$ &$0.916$\\  
$CI_{boot} $ &    $0.944$  &$0.939$ &$0.920$\\
$CI_{u}$ & $0.948$ & $0.943$ &$0.928$\\
$CI_{p}$  & $0.943$ & $0.947$ &$0.919$\\
$CI_{t}$  &       $0.949$ & $0.941$ &$0.925$\\
\hline
\end{tabular}			
\end{center}			
\end{table}
\begin{table}	
\caption {Average coverage rates for several confidence intervals in \barma$(1,1)$ model.}
\tablesizea
\label{T:barma2}
\begin{center}
\begin{tabular}{rrrrr}
\hline
Intervals 	&	${\alpha}$  &  ${\varphi}_{1}$ &  ${\theta}_{1}$ & ${\phi}$ \\        
\hline 
\multicolumn{5}{c}{$n=20$}\\
\hline                
$IC$  &    $0.770$ &$0.762$ & $0.690$ &$0.936$\\  
$CI_{boot} $  &   $0.867$ &$0.880$ & $0.884$ &$0.960$\\
$CI_{u}$ & $ 0.810$ &$0.807$ & $0.816$ &$0.908$\\
$CI_{p}$ & $0.888$ &$0.912$ & $0.935$ &$0.923$\\
$CI_{t}$   &  $ 0.881$&$ 0.900$ & $0.903$ &$0.967$\\               
\hline
\multicolumn{5}{c}{$n=30$}\\
\hline                
$IC$  &  $0.718$ & $0.710$ &  $0.688$ &  $0.943$ \\  
$CI_{boot} $  &   $0.870$ & $0.877$ &  $0.884$ & $0.961$\\
$CI_{u}$ & $0.813$ & $0.814$ &  $0.857$ & $0.936$\\
$CI_{p}$ &$0.921$ & $0.932 $ & $0.939$ &  $0.909$ \\
$CI_{t}$   & $0.881$ & $ 0.888$ &   $0.899$ &  $0.967$ \\                
\hline
\multicolumn{5}{c}{$n=50$}\\                           
\hline
$IC$  &         $0.718$ &$0.708$ & $0.715$ &$0.932$\\
$CI_{boot} $  &      $0.905$ &$0.902$ &$ 0.925$ &$0.947$\\
$CI_{u}$ &  $0.864$ &$0.853$ & $0.909$ &$0.936$\\
$CI_{p}$ & $0.930$ &$0.929$ & $0.947$ &$0.933$\\
$CI_{t}$   &       $  0.910$ &$0.910$ & $0.931$ &$0.952$\\
\hline
\multicolumn{5}{c}{$n=100$}\\                           
\hline
$IC$  &        $ 0.704$ &$0.705$ & $0.718$ &$0.942$\\
$CI_{boot} $  &      $0.927$& $0.924$ & $0.934$ &$0.951$\\
$CI_{u}$ &  $0.877$ &$0.877$ & $0.907$ &$0.945$\\
$CI_{p}$ &$0.928$ &$0.923$ & $0.889$ &$0.949$\\
$CI_{t}$   &       $ 0.928$ &$0.929$ & $0.934$ &$0.955$\\
\hline
\end{tabular}			
\end{center}			
\end{table}

\end{document}